\documentclass[fleqn,usenatbib]{mnras}

\usepackage{newtxtext,newtxmath,graphicx,amsmath,amssymb,comment}

\usepackage[T1]{fontenc}
\usepackage{ae,aecompl}

\usepackage{graphicx}	
\usepackage{amsmath}	
\usepackage{amssymb}	

\usepackage{xcolor}

\newcommand{\HI}{H\,\textsc{i}}

\title[Radio Loud Black Holes at Cosmic Dawn]{The Radio Scream from Black Holes at Cosmic Dawn: A Semi-Analytic Model for the Impact of Radio Loud Black-Holes on the 21\,cm Global Signal}

\author[A. Ewall-Wice et al.]{
Aaron Ewall-Wice,$^{1,2,3}$\thanks{E-mail: aaronew@berkeley.edu}, Tzu-Ching Chang$^{1,4}$, T.~Joseph W.~Lazio$^{1}$
\\
$^{1}$Jet Propulsion Laboratory, California Institute of Technology 4800 Oak Grove Dr, M/S 169-237, Pasadena CA 91109, USA\\
$^{2}$Department of Astronomy, UC Berkeley, Berkeley, CA 94720 USA \\
$^{3}$Berkeley Center for Cosmological Physics, UC Berkeley, Berkeley, CA 94720, USA\\
$^{4}$California Institute of Technology, 1200 E California Blvd, Pasadena, CA 91125, USA\\
}

\date{Accepted XXX. Received YYY; in original form ZZZ}

\pubyear{2019}
\begin{document}
\label{firstpage}
\pagerange{\pageref{firstpage}--\pageref{lastpage}}
\maketitle

\begin{abstract}
We use a semi-analytic model to explore the potential impact of a brief and violent period of radio-loud accretion onto black-holes ({\it The Radio Scream}) during the Cosmic Dawn on the \ion{H}{1} hyperfine 21\,cm signal. We find that radio emission from super-massive black hole seeds can impact the global 21\,cm signal at the level of tens to hundreds of percent provided that they were as radio loud as $z\approx1$ black holes and obscured by gas with column depths of  $N_\text{H}\gtrsim 10^{23}$\,cm$^{-2}$. We determine plausible sets of parameters that reproduce some of the striking features of the EDGES absorption feature including its depth, timing, and side steepness while producing radio/X-ray backgrounds and source counts that are consistent with published limits. Scenarios yielding a dramatic 21\,cm signature also predict large populations of $\sim \mu$Jy point sources that will be detectable in future deep surveys from the Square Kilometer Array (SKA). Thus, 21\,cm measurements, complemented by deep point source surveys, have the potential to constrain optimistic scenarios where super-massive black-hole progenitors were radio-loud.
\end{abstract}

\begin{keywords}
cosmology: dark ages, reionization, first stars -- general:radio continuum -- quasars: general -- galaxies: Intergalactic medium -- galaxies: high-redshift
\end{keywords}



\section{Introduction}
Before stars and quasars established strong ionizing backgrounds, most of the Hydrogen in the Universe existed as \HI\ in the intergalactic medium (IGM). Since \HI's hyperfine transition is sensitive to astrophysical backgrounds, its emitted 21\,cm photons carry considerable information on the first sources of radiation such as stars, supernovae, black holes, and dark matter.  \citep[For reviews, see][]{FurlanettoBriggsOh:2006,Morales:2010,Pritchard:2012,McQuinn:2016})

A long-standing question that 21\,cm observations might help resolve is how $\gtrsim 10^9 \text{M}_\odot$ super-massive black holes (SMBH) observed at $z \sim 7$ \citep{Mortlock:2011,Wu:2015,Banados:2017} could assemble from much smaller seeds in less than a billion years.  Several studies have investigated how 21\,cm observations might reveal the  heating and ionization of H\,\textsc{i} in the intergalactic medium (IGM)
due to the X-ray or ultraviolet (UV) emission from accreting black-hole seeds
\citep{Madau:2004,Zaroubi:2007,Ripamonti:2008,Madau:2015,Tanaka:2016}. Radio emission, a well-known product of accretion on to black holes, has received almost no consideration. 

This lack of attention is reasonable. It has long been recognized that the diffuse lobes, that dominate radio emission in many low-redshift active galactic nuclei (AGN), should be suppressed at high redshift. The higher energy density of the Cosmic Microwave Background (CMB) is expected to cause relativistic electrons to shed their energy through inverse Compton (IC) scattering instead of synchrotron emission (SE) (e.g., \citealt{Ghisellini:2014,Ghisellini:2015,Saxena:2017}). IC suppression of SE may partially explain the dearth of SDSS quasars matched with radio sources at $z\gtrsim 3$ \citep{Haiman:2004,McGreer:2009,Volonteri:2011}. In addition, some studies \citep[e.g.,][]{Jiang:2007} claim tentative detections of decreasing radio loud source densities at high redshift.\footnote{While its specific definition varies throughout the literature, we define ``radio-loudness'' as the log$_{10}$-ratio between emission frame 2.8\,GHz luminosity and 4000\,\AA  luminosities. } 

At the same time, observations of the highest redshift radio sources by \citet{Banados:2015} find no evidence for a decrease in the radio-loud fraction while \citet{Ghisellini:2016} find that the lack of radio-loud SDSS quasars relative to known blazar counts might be heavily influenced by obscuration. It is entirely possible that the first black-hole seeds had high spins, large magnetic fields, and accreted in dense environments that could allow the relativistic electrons in their jets to have emitted the bulk of their energy through radio SE, despite the higher CMB energy densities. The nature and evolution of radio emission from the highest redshift black-holes remain essentially unconstrained by direct measurements.

The question as to whether black-holes between $10 \lesssim z \lesssim 30$ generated appreciable radio emission is made more compelling by tentative observations by the ARCADE-2 \citep{Fixsen:2011} and EDGES \citep{Bowman:2018} (B18) experiments. The first, made at $\sim$ GHz frequencies, observes a radio monopole that exceeds the level expected from known populations of extragalactic radio sources. Recent observations by the Long Wavelength Array between 40 and 80\,MHz lend further support for the existence of such a background \citep{Dowell:2018}, though it is still unclear whether it arises from lower redshifts or is purely a result of observational systematics \citep{Singal:2018}. The EDGES feature, at $\sim 70$\,MHz reports a 21\,cm absorption trough with an amplitude that greatly exceeds what is possible for adiabatically cooled H\,\textsc{i} gas absorbing solely the CMB background.

One possible explanation for the unusual timing and depth of the EDGES feature might be an enhanced radio background, similar to what was reported by ARCADE-2 and originating at $z \gtrsim 17$ \citep{Feng:2018,Fialkov:2019}. If confirmed, this radio background might be sourced by black-holes \citep{EwallWice:2018}, star-forming galaxies (SFGs) \citep{Mirocha:2018}, annihilations of an axion-like dark-matter particle \citep{Fraser:2018} or dark photons \citep{Pospelov:2018}. \citealt{Sharma:2018} (S18) demonstrated that astrophysical sources of a CD radio background require $\sim$\,mG magnetic fields. They also argue that even $\gtrsim$\,mG sources would need to be $\sim 1000$ times more radio loud than sources at $z\sim 0$. While there are significant challenges for sources maintaining $\sim$\,mG magnetic fields over large time-scales, we disagree with S18's latter argument and think that radio loudness at the level of today's sources is sufficient. We explain why in \S~\ref{ssec:IC}. 

If early accreting black-holes could maintain $\sim$\,mG magnetic fields and produce appreciable radio emission, they might have an observable effect on the 21\,cm signal. Should this be the case, a self consistent model for black-hole radio emission, along with other mechanisms for radio emission (such as SFGs, axions, or dark photons) will be necessary to interpret 21\,cm observations. Such models will also be needed to distinguish a radio-background explanation for EDGES from alternative theories such as dark matter cooling \citep{Barkana:2018a,Barkana:2018b,Munoz:2018,Fialkov:2018,Berlin:2018} or systematics \citep{Hills:2018, Sims:2019, Singh:2019}. Radio background models can also be used to predict discrete radio source populations that might be detected in surveys. 

In a previous paper, \citet{EwallWice:2018}, we estimated the amplitude of the radio background that might arise from vigorously accreting black-hole seeds at high redshift and discussed the potential for such a radio background to explain EDGES assuming saturated Ly~$\alpha$ coupling of the spin-temperature, $T_s$, to the gas kinetic temperature, $T_k$.  We ignored X-ray heating from the same sources. Our conclusions in this paper were overly generous since heating and incomplete Ly~$\alpha$ coupling can both reduce the amplitude of the Cosmic Dawn (CD) absorption signal. The logical next step in any modeling effort is to self-consistently compute the 21\,cm signal under the influence of all radiative outputs from accreting black holes and star formation; including the effects of radio, X-ray heating, Ly~$\alpha$ coupling, and UV/X-ray ionization.

In this follow-up study, we construct a semi-analytic model to predict the evolution of the global 21\,cm signal under the influence of accreting black-holes radiating across the electromagnetic spectrum; from radio waves to X-rays. We use this model to determine how radio emission from the first black-holes might appear in global 21\,cm observations. We also explore whether these radio-loud black-holes might explain the EDGES feature and what scenarios might be detected in current and up-coming radio surveys. 

This paper is organized as follows. We begin with an overview of the existing global-signal framework in \S~\ref{sec:OldModel} before describing the modifications introduced by black-hole seeds in \S~\ref{sec:NewModel}. We explore the impact of radio-loud accretion on the global signal in \S~\ref{sec:ModelingResults} by computing the global-signal for a variety of illustrative models. We discuss our perceived flaws in S18's argument for the complete impossibility of radio loud sources during the CD along with the significant challenges that still remain in \S~\ref{ssec:IC}. We explore the degeneracies in several models that yield an absorption feature similar to the one reported by EDGES in \S~\ref{ssec:EDGES}, how they might be broken by future point-source surveys, and discuss which features we are and are not able to reproduce. We conclude in \S~\ref{sec:Conclusion}.


\section{The Modeling Framework}\label{sec:OldModel}

Luminous sources affect the physical properties of H\,\textsc{i} in the IGM by setting up radiative backgrounds. Throughout this paper, we assume that the co-moving emissivity of radiation, with frequency $\nu$, at position ${\bf x}$ and redshift $z$ is the sum of contributions from stars/stellar remnants and AGN, which we denote as $\epsilon_\star$ and $\epsilon_\bullet$ respectively. 
\begin{equation}
\epsilon({\bf x},z,\nu) = \epsilon_\star({\bf x},z,\nu) + \epsilon_\bullet({\bf x},z,\nu).
\end{equation}

In this section, we describe the existing framework for computing the global 21\,cm signal and previously derived expressions for $\epsilon_\star$. In the next section we will discuss how we modify this framework by adding accreting radio-loud black-holes.
We will start out with a review of the existing frameworks for computing $\epsilon_\star$ for Ly~$\alpha$ (\S~\ref{ssec:LyAlpha}), X-ray (\S~\ref{ssec:Xrays}), and UV continuum (\S~\ref{ssec:UV}) photons. We end this section with a description of how radiative backgrounds impact the observable 21\,cm brightness temperature, $\delta T_b$ (\S~\ref{ssec:dTb}).  

In this work, we chose to focus on AGN and not consider the radio emission from X-ray binaries (HMXB). In the local Universe, X-ray binaries are roughly four orders of magnitude less radio loud than AGN \citep{Heinz:2003} and their total contribution to the radio background is negligible compared to SFGs and AGN. In addition, should such sources exist in large enough abundance to produce a large radio background, obscuring them could be difficult. A study by \citet{Das:2017} finds that most XRBs in the early universe would have column depths too small to prevent $\approx 1$\,keV X-rays from significantly heating intergalactic gas. 

On the other hand, a substantial radio background from HMXB could arise provided that the active fraction of binary black holes was much larger then observed today \citep{Mirabel:2019}. Furthermore, the \citet{Das:2017} study investigated the column densities to star formation in $z \sim 7$ atomic cooling halos whereas Pop-III HMXBs might preferentially form in more obscured environments. 

\subsection{Ly~$\alpha$ Emission}\label{ssec:LyAlpha}
Ly~$\alpha$ photons primarily impact the 21\,cm signal by coupling the  H\,\textsc{i} spin temperature, $T_s$, to its kinetic temperature $T_k$ through a multiplicative coupling constant, $x_\alpha$ (see equation~\ref{eq:Ts} below).
As outlined in \citet{Hirata:2006}, whose recipe we employ here, $x_\alpha$ as proportional to the Ly~$\alpha$ {\it number} flux, $J_\alpha$. $J_\alpha$ includes a contributions from secondary photons stimulated by X-rays, ($J_{\alpha,X}$) and a contribution from atoms excited by UV continuum photons that redshift into the Ly~$n\ge2$ transitions ($J_{\alpha,\text{UV}}$). We compute $J_{\alpha,\text{UV}}$ using the prescription in \citet{Hirata:2006}. 
\begin{equation}
J_{\alpha,\text{UV}}(z)=\sum_{n=2}f_\text{rec}(n)\frac{c(1+z)^2}{4 \pi } \int_z^{z_\text{max}(n)}\frac{\epsilon_\text{UV}[\nu_n(z,z'),z']}{H(z')h_P \nu_n(z,z')} dz', 
\end{equation}
where $h_P$ is Planck's constant, $\nu_n(z,z^\prime)$ is the emitted frequency of a photon at $z^\prime$ that redshifts into the Ly~$n$ resonance at redshift $z$,  $f_\text{rec}(n)$ is the probability of an absorbed Ly~$n$ photon being re-emitted as a Ly~$\alpha$ photon, and $z_\text{max}(z,n)$ is the maximum redshift from which a Ly~$n$ photon can be emitted without redshifting into Ly~$(n-1)$ and being absorbed before reaching redshift $z$.

We write the co-moving emissivity of UV photons with frequency $\nu$ from redshift~$z$ as $\epsilon_\text{UV}(\nu,z)$.  We split $\epsilon_\text{UV}$ into contributions from stars, $\epsilon_{\text{UV}\star}$, and black holes, $\epsilon_{\text{UV}\bullet}$. Here we focus on the stellar contribution and will discuss the black-hole contribution in \S~\ref{sssec:Optical}. We set $\epsilon_{\text{UV}\star}$ to be proportional to the star formation rate density.

\begin{equation}
\epsilon_{\text{UV}\star}(\nu,z)=h_P \nu f_\star  N_\gamma n_\text{UV}(\nu)\frac{\Omega_b}{\Omega_m}\frac{\dot{\rho}_{\text{coll}\star}(z)}{\mu m_p},
\end{equation}

where $f_\star$ is the fraction of baryons incorporated into stars, $\mu$ is the reduced particle mass of the IGM, $m_p$ is the proton mass, $N_\gamma$ is the average number of ionizing photons emitted per stellar baryon, and $n_\text{UV}(\nu)$ is the differential number of photons emitted per unit frequency divided by $N_\gamma$ calculated in  \citet{Barkana:2005}.  
The quantity $\rho_{\text{coll}\star}$ is the {\it total} comoving density of matter collapsed in halos with virial-temperatures above a minimum threshold, $\text{T}_{\text{vir}\star}^\text{min}$ computed from the \citet{Sheth:1999} mass function.  The values for $f_\star$ and $N_\gamma$ are currently unknown. As fiducial values, we adopt  $f_\star=0.1$ and $N_\gamma=2000$, which yields an ionization history that is consistent with the  joint constraints derived from measurements of the CMB optical depth, quasar absorption features, and the kinetic Sunyaev Zeldovich effect \citep{Greig:2017}. 

\subsection{X-rays}\label{ssec:Xrays}
X-rays from early galaxies are thought to have had a significant impact on the physical state of H\,\textsc{i} during the Cosmic Dawn, primarily by heating the gas (raising $T_k$) but also by ionizing it (raising the electron fraction, $x_e$) and stimulating secondary UV photons. At low redshifts, X-ray emission from SFGs is contributed to by two types of sources. First are high-mass X-ray binaries and hot ISM generated by supernovae \citep{Pacucci:2014} whose emissivities are proportional to star formation rate. Secondly, low-mass X-ray binaries contribute emission that traces the total star formation history. At higher redshifts, high mass X-ray binaries and ISM dominate. We therefore adopt an X-ray emissivity based on the empirical X-ray luminosity star formation rate density relation that appears in much of the 21\,cm literature (e.g. \citet{Oh:2001}, \citet{Mirabel:2011}, \citet{Fialkov:2014}, \citet{Mesinger:2013}). 

\begin{align}
\epsilon_{X\star}(E) =& 3 \times 10^{39} \text{erg sec}^{-1} \text{keV}^{-1} f_X\,\left(h^2 \text{M}_\odot \text{yr}^{-1} \text{Mpc}^{-3}\right)^{-1}  \frac{\Omega_b}{\Omega_m} \nonumber \\ & \times A_{2,10}(\alpha_{X\star}) 
f_\star \dot{\rho}_{\text{coll}\star}(z)\left(\frac{E}{\text{keV}}\right)^{-\alpha_{X\star}} \nonumber \\
& \times \exp\left\{-N_{\text{H}\star}\left[ \sigma_\text{H}(E)+\chi \sigma_\text{HeI}(E)\right] \right\} ,
\end{align}

where $N_{\text{H}\star}$ is the Hydrogen column density to the X-ray sources, $\sigma_\text{H}(E)$ and $\sigma_\text{HeI}(E)$ are the Hydrogen and Helium photo-ionization cross sections, $\chi$ is the ratio between Helium and Hydrogen number densities\footnote{$\chi=Y_P/4/(1-Y_P)$ where $Y_P$ is the cosmological Helium mass fraction.}, and $A_{2,10}(\alpha_X) \equiv (1-\alpha_X)(10^{1-\alpha_X}-2^{1-\alpha_X})^{-1}$ is a normalization factor ensuring that when $f_X=1$, the 2-10\,keV emissivity matches observations of local SFGs by \citet{Mineo:2012}. We compute X-ray ionization, heating, and the generation of secondary Ly~$\alpha$ photons using standard radiative transfer recipes using interpolation of the numerical cross sections from \citet{Furlanetto:2010}\footnote{for example, see M11}. To compute $I_X$ at redshift $z$, we integrate the contributions of sources at higher redshifts and evolve the kinetic temperature and X-ray ionization fractions using equations 8 and 9 in \citet{Mesinger:2011} (M11). 
We adopt fiducial values of $f_X=1$, $\alpha_{X\star}=1.2$, and $N_{\text{H}\star}=10^{21}$\,cm$^{-2}$ which are typical values obtained in numerical simulations of early galaxies \citep{Das:2017}.

\subsection{Ionizing UV Continuum Photons}\label{ssec:UV}
The majority of Hydrogen reionization was probably driven by UV photons generated by star formation and/or AGN activity. We track the evolution of the volumetric fraction of H\,\textsc{ii} regions~$x_i$ through the differential equation, 
\begin{equation}\label{eq:StellarIonization}
\dot{x}_i = \frac{4}{4-3 Y_p} f_{\text{esc}\star}f_\star N_\gamma \frac{\dot{\rho}_{\text{coll}\star}(z)}{\rho_0}-n_{H,0} x_i C(z)\alpha_A(T_4) (1+z)^3,
\end{equation}
where $f_{\text{esc}\star}$ is the escape-fraction for stellar UV photons; $n_{H,0}$ is the comoving number density of Hydrogen atoms; $\alpha_A$ is the case~A recombination coefficient; $T_4$ is the electron temperature in  H\,\textsc{ii} regions, which we take to be $10^4$\,K; and $C(z)$ is the clumping factor, which we set equal to $C(z)=2.9\times((1+z)/6)^{-1.1}$ \citep{Madau:2015}. We choose $f_{\text{esc}\star}=0.1$ as a fiducial value. We note that the $x_i$ in equation~\ref{eq:StellarIonization} refers to H~\textsc{ii} regions generated by UV photons. Ionizations from X-rays are handled separately by evolving $x_e$ using equations 8 and 9 in M11.

\subsection{The 21\,cm Brightness Temperature}\label{ssec:dTb}
The 21\,cm brightness temperature $\delta T_b({\bf x})$ is a function of position while the ``global" signal is equal to the ensemble average, $\langle \delta T_b({\bf x}) \rangle$. As we will only be discussing the average ``global'' 21\,cm signal, we let $\delta T_b \equiv \langle \delta T_b ({\bf x})\rangle$ .  We approximate $\delta T_b$ using the equation\footnote{For our computation of the global signal, the terms in equation~\ref{eq:Tb} are their volume averaged quantities which ignores the spatial correlations that exist between $x_{HI}$, $T_s$, $T_r$, and the density field. \citet{Liu:2016} find that this approximation has a $\sim 10\,\%$ impact on the global signal amplitude though their analysis ignored $T_s$.}

\begin{equation}
\label{eq:Tb}
\delta T_b \approx 27\, x_\text{HI} \left(\frac{\Omega_b h^2}{0.023}\right) \left(\frac{0.15}{\Omega_m h^2} \right) \left( 1-\frac{T_r}{T_s} \right)\sqrt{\frac{1+z}{10}}\,\text{mK}
\end{equation}

where $T_r$ is the average brightness temperature of the 21\,cm radio background at redshift~$z$, $T_s$ is the H\,\textsc{i} spin-temperature, $x_\text{HI}$ is the average neutral fraction, $\Omega_b$ and $\Omega_m$ are the fractions of the critical energy density in baryons and matter respectively, and $h=H_0/100$\,km\,sec$^{-1}$ Mpc$^{-1}$ where $H_0$ is the Hubble constant. $T_r$ is typically set to be equal to the temperature of the CMB, $T_r = T_\text{CMB}$. In this paper, we consider an additional contribution to $T_r$ from \hbox{AGN}.  The coupling of H\,\textsc{i}'s spin temperature~$T_s$, to its kinetic temperature~$T_k$, $T_r$, and the Ly~$\alpha$ color temperature~$T_\alpha$, is described by (\citealt{Field:1958})

\begin{equation}\label{eq:Ts}
T_s^{-1} = \frac{T_r^{-1}+ x_\alpha T_\alpha^{-1} + x_c T_k^{-1}}{1+x_\alpha+x_c},
\end{equation}

where $x_\alpha$ and $x_c$ are the Ly~$\alpha$ and collisional coupling constants.  The factor $x_\alpha$ is determined by the Ly~$\alpha$ flux relative to the radiative background while $x_c$ is determined by the density, temperature, and ionization state of H\,\textsc{i} (\citealt{FurlanettoBriggsOh:2006}). 
In this work, the radio temperature, $T_r$ arises from the CMB and the radio background from black-holes.

\begin{align}
T_r(\nu,z) = & T_\text{CMB}(z) \nonumber \\
           + & \frac{\lambda_{21}^2}{2 k_B} \frac{c(1+z)^3}{4\pi} \int \epsilon_{\text{R}\bullet}\left[\nu\frac{1+z^\prime}{1+z},z^\prime \right](1+z^\prime)^{-1}H^{-1}(z^\prime) dz^\prime,
\end{align}

where $\epsilon_{\text{R}\bullet}(\nu,z)$ is the co-moving radio emissivity of black-holes at frequency $\nu$ and redshift $z$ which we discuss in detail in \S~\ref{ssec:RadioLoudness}, $k_B$ is the Boltzmann constant, $c$ is the speed of light in vacuum, and $H(z)$ is the Hubble parameter at redshift $z$. 
The Lyman-$\alpha$ color temperature is computed from $T_\alpha(z) =  \frac{\lambda_\alpha^2}{ 2 k_B } I_\alpha(z)$ where $I_\alpha(z)$ is the background intensity of Lyman-$\alpha$ photons frome the de-excitation of Ly~$n$ states and secondary production by X-rays (\S~\ref{ssec:LyAlpha}). 

In addition to the heating from X-rays (\S~\ref{ssec:Xrays}), we also include the impacts of adiabatic cooling, Compton heating, and ionization induced changes in thermal degrees of freedom when evolving $T_k$. 

While X-rays are expected to the primary drivers of IGM heating, recent work by \citet{Venumadhav:2018} (V18) finds that a radio-background can have an significant impact on $T_k$ when the IGM is unheated. We include X-rays in this work and also find that some heating has occurred at the absorption minimum in all of our models. We therefore choose to ignore the V18 CMB heating mechanism in this paper. 

\section{Modeling the Black-Holes}\label{sec:NewModel}
We now discuss how we add the impact of radiation from exponentially growing black-hole seeds to the global signal framework described in \S~\ref{sec:OldModel}. We start with a simple equation that allows us to compute the black-hole density as a function of redshift (\S~\ref{ssec:BHDensity}). Assuming that our black-holes grow and radiate through Eddington limited accretion allows us to compute their comoving emissivity across the electromagnetic spectrum \S~\ref{ssec:BHRadiation}. 

\subsection{Evolving the Black Hole Density Field}\label{ssec:BHDensity}

\begin{figure}
\includegraphics[width=.5\textwidth]{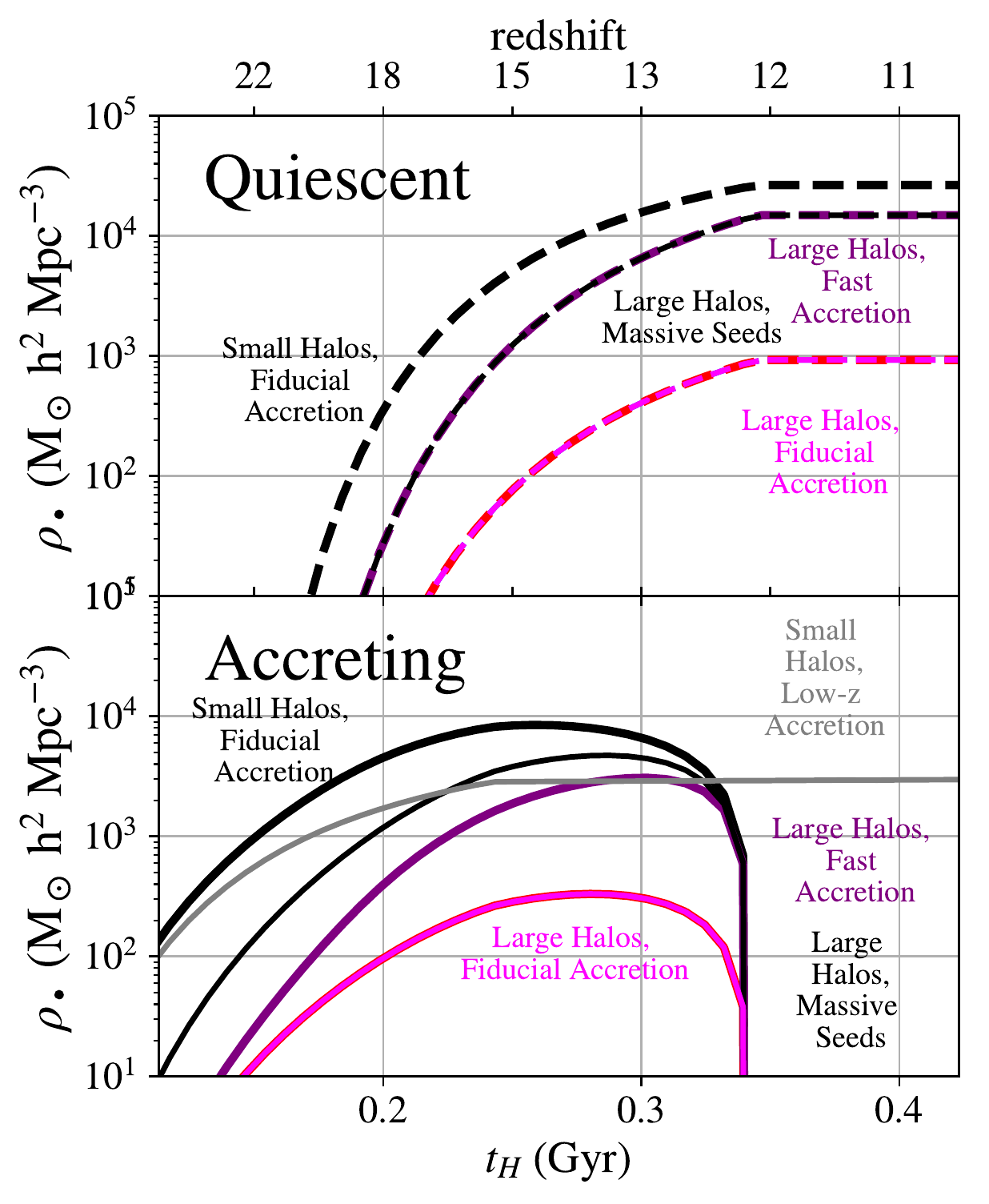}
\caption{The evolution of the quiescent black-hole density~$\rho_{\bullet \text{q}}$ (top panel) and the accreting black-hole density~$\rho_{\bullet \text{a}}$ (bottom panel). Seed formation cuts off at $z_\text{min}^i=16$ for all of our models, causing $\rho_{\bullet\text{a}}$ to level off. With an accretion lifetime of $100$\,Myr, $\rho_{\bullet\text{a}}$ drops to zero 100 Myr later at $z \approx 12$. 
Lower halo mass thresholds and/or higher mass seeds translate the onset of density growth to earlier times while smaller black-hole mass e-folding (Salpeter) times increase its steepness. Note that in our model, the number density, not overall mass, of halos determines the black-hole density. Thus Large Halos scenarios have fewer black holes (and overall lower average black hole density) than the Small Halos model.}\label{fig:Densities}
\end{figure}

 Fig.~\ref{fig:Densities} illustrates several density evolution scenarios for accreting ($\rho_{\bullet\text{a}}$) and quiescent ($\rho_{\bullet\text{q}}$) black holes in our models. The remainder of this section describes our derivation of these histories. 

In our framework, black-hole seeds form with a uniform mass $m_{\bullet}^i$ in a fraction $f_\bullet$ of newly collapsed halos up to a time $t^i_{\text{max}}$ (redshift $z^i_{\text{min}}$). We use the superscript ``i'' to denote that the masses are ``initial''. 

Once a black-hole forms, we count its mass as contributing to $\rho_{\bullet\text{a}}$ and let it grow exponentially with an e-folding (Salpeter) time of $\tau_{\text{s}}$. After an active lifetime of $\tau_\text{L}$, we assume that each black-hole stops growing/shining as a result of feedback and/or exhausting its fuel supply. At this point, we add its mass to $\rho_{\bullet\text{q}}$.

With these assumptions, we write down integral equations governing the evolution of $\rho_{\bullet\text{q}}$ and $\rho_{\bullet\text{a}}$ that we integrate numerically. 

\begin{align}\label{eq:BlackHoleGrowthAccretion}
\rho_{\bullet a}(t) = m_\bullet^i \int_0^{\tau_\text{L}}  \dot{n}_\bullet \left[ t - t^\prime \right] e^{t^\prime/\tau_\text{s}} d t^\prime,
\end{align}

where

\begin{equation}
\dot{n}_\bullet(t) = f_\bullet \begin{cases} \frac{d}{dt}\int_{m_\text{min}}^{m_\text{max}} n_h(m,t) dm  & t \le t_\text{max}^i \\
0 & \text{otherwise}
\end{cases}
\end{equation}

and $n_h(m,t) dm$ is the comoving number density of dark-matter halos with masses between $m$ and $m+dm$ at time $t$. The quiescent black hole density is given by

\begin{align}\label{eq:BlackHoleGrowthQuiescent}
\dot{\rho}_{\bullet\text{q}}(t)= m_\bullet^i e^{\tau_\text{L}/\tau_\text{s}} \int_0^t \dot{n}_\bullet\left( t^\prime \right) dt^\prime
\end{align}

While we choose to integrate in the time domain, our recipe for density evolution is functionally the same as the one used by \citet{Yue:2013} to predict the contribution of black-holes to the high redshift infrared and X-ray backgrounds. 

For our fiducial model, we consider a scenario in which black-holes form from massive Pop-III stars collapsing directly into black holes with $m_\bullet^i = 10^2 \text{M}_\odot$ in $10\%$ of halos \citep{Hirano:2015} with masses between $T_\text{vir}^\text{min} = 2000$K and $T_\text{vir}^\text{max} = 10^4$\,K (halo masses of $ m \approx 1.3 \times 10^6 \text{M}_\odot h^{-1}$ and $m \approx 1.5 \times 10^7 \text{M}_\odot h^{-1}$ at $z=12$ respectively). 
We adopt a fiducial Salpeter time of 
\begin{align}\label{eq:taccrete}
\tau_{\text{s}}&=\frac{\tau_{\text{E}} \eta }{\lambda f_{\text{duty}}}\nonumber \\
& \approx 45 \left(\frac{\eta}{0.05}\right) \left(\frac{1}{\lambda}\right)\left(\frac{0.5}{f_{\text{duty}}}\right)\text{Myr}
\end{align}
where $\tau_{\text{E}}$ is the Eddington time-scale of $0.45$\,Gyr, $\eta$ is the fraction of infalling rest-mass that is emitted as radiation, $\lambda$ is the fraction of the Eddington luminosity that the black-hole radiates, and $f_{\text{duty}}$ is the accretion duty cycle.
 
 Accretion rates for black-hole seeds at $z \gtrsim 10$ are, so-far, unknown. We settle on a fiducial value for $\tau_\text{s}$ of $45$\,Myr, corresponding radiative efficiencies in the range of $0.025-0.1$ \citep{Shankar:2010} and time-averaged Eddington ratios between $0.1$ and $10$, similar to what is observed at low redshift \citep{Willott:2010}. We also employ a fiducial lifetime of $\tau_\text{L} = 100$\,Myr; which is at the upper end of inferred accretion lifetimes observed at low-redshift \citep{Bird:2008,Shabala:2008,Turner:2015} but is consistent with accretion lifetimes in simulations of intermediate mass black-hole seeds growing in dense environments \citep{Pacucci:2015}. We choose a fiducial minimum redshift for seed formation of $z_{\text{seed}}^i=16$, which is within the range that Lyman-Werner regulated Pop-III formation has been found to cut off in semi-analytic studies \citep{Mebane:2017}. Note that while seed formation ends at $z_\text{min}^i = 16$, black holes continue to emit down to $z \approx 12$ for our choice of parameters (Fig.~\ref{fig:Densities}).
 While we use the above ``fiducial'' values as reference points, our goal is to understand how specifically the signal changes when we vary them. 
We list our model's parameters for black-hole growth along with their fiducial values in Table~\ref{tab:MassParams}. 

In order to build some intuition on equations~\ref{eq:BlackHoleGrowthAccretion} and \ref{eq:BlackHoleGrowthQuiescent}, it is useful to inspect some general patterns in the density histories evolved with these equations (Fig.~\ref{fig:Densities}). ``Small halos, fiducial accretion'' corresponds to our fiducial model. We will discuss the specific deviations between each model in \S~\ref{ssec:Models}. At early times, when $t \lesssim \tau_\text{s}$, $\rho_{\bullet\text{a}}$ grows proportional to $\dot{\rho}_\text{coll}$ but quickly exceeds this growth after $t \gtrsim \tau_\text{s}$. New seeds stop forming after $z_\text{min}^i$, causing $\rho_{\bullet\text{a}}$ to turn over and fall to zero after $t_\text{max}^i + \tau_\text{L}$. The `small halos, low-z' line plateaus after $z_\text{min}^i$ but this is simply because the Salpeter time and accretion lifetime (1\,Gyr and 4.5\,Gyr respectively) are much larger than the time interval on the plot. Quiescent densities remain at zero until $\tau_\text{L}$ at which point they rapidly grow and flatten out at $t_\text{max}^i + \tau_\text{L}$ as the last accreting black holes die out and stop feeding the quiescent population. 

In this work, we focus on the average black-hole density and its impact on the global 21\,cm signal. However, our formalism can also be adopted into simulations of 21\,cm fluctuations. One would simply track the black-hole density as a function of position by integrating equations \ref{eq:BlackHoleGrowthAccretion} and \ref{eq:BlackHoleGrowthQuiescent} in each voxel using the spatially varying  $\dot{\rho}_{\text{coll}}$ instead of the global value, similar to what is done in the semi-numerical models {\tt 21cmFAST} (M11) and {\tt simfast21} \citep{Santos:2008}.

\begin{table}
\begin{tabular}{|c|l|c|}
Parameter & Description & Fiducial Value \\ \hline
$\tau_\text{s}$ &  black-hole mass $e$-folding time (Salpeter) & 45\,Myr \\
$\tau_\text{L}$ & accretion lifetime & 100\,Myr \\
$z_\text{min}^i$ &  minimum redshift for & \\ & black-hole seed formation & $16$ \\
$T_{\text{vir}\bullet}^\text{min}$ & minimum virial temperature & \\ & of black-hole seed halos & $2000$\,K \\
$T_{\text{vir}\bullet}^\text{max}$ & maximum virial temperature & \\ & of black-hole seed halos &
$10^4$\,K \\
$m_\bullet^i$ & black-hole seed mass  & $100$\,M$_\odot$ \\
$f_\bullet$ & fraction of halos with seeds & $10^{-1}$
\end{tabular}
\caption{Parameters associated with black-hole mass evolution and their ``fiducial" values.}\label{tab:MassParams}
\end{table}



\subsection{Radiation Backgrounds from Black Holes}\label{ssec:BHRadiation}
 We now describe our prescription for computing the radiative backgrounds from the black-hole mass density. These backgrounds include X-rays (\S~\ref{sssec:Xrays}), radio waves (\S~\ref{sssec:Radio}), and UV photons (\S~\ref{sssec:Optical}).

\subsubsection{X-rays}\label{sssec:Xrays}
%
X-ray emission from black-holes originates from three sources associated with accretion: thermal emission from the accretion disk, IC scattering of accretion disk photons off of hot electrons in the corona, and IC scattering of CMB photons by relativistic jet electrons. 

Studies typically quantify the total X-ray luminosity of an
AGN with the parameter $k_\text{bol} \equiv L_\text{bol}/L_{[2-10]\text{keV}}$. At low redshift, \citet{Lusso:2010,Marchese:2012} observe $k_\text{bol}^{-1}\approx 0.06$ with X-ray emission dominated by IC upscattering of soft photons in the corona. AGN X-ray spectra usually exhibit an absorbed power-law with a spectral index of $\alpha_X \approx 0.9$ \citep{Nandra:1994} and a high-energy exponential cutoff at $300$\,keV \citep{Titarchuk:1994}.

 We set the amplitude of the time-averaged black-hole X-ray emissivity from the disk and corona through the accretion rate and a normalization parameter $g_\text{bol}$. $\int dE \epsilon_X(E) \propto g_\text{bol} \tau_\text{s}^{-1} \rho_{\bullet\text{a}}$, where 
\begin{equation}\label{eq:GBOL}
g_\text{bol} \equiv 0.003 \left(\frac{\eta}{0.05}\right)\left(\frac{k_{\text{bol}}^{-1}}{0.06}\right).
\end{equation}
We have included $\eta$ in our definition of $g_\text{bol}$ to cancel out the factor of $\eta$ in $\tau_\text{s}$ when these two variables are multiplied together. 


 Multiplying the bolometric accretion factor by an absorbed power-law and an exponential cutoff gives us our expression for the X-ray emissivity of black-holes, 

\begin{align}\label{eq:XrayEmissivity}
\epsilon_{\text{X}\bullet}(E) \approx& 2 \times 10^{49} \left(\frac{45\,\text{Myr}}{\tau_\text{s}}\right) \nonumber \\ \times & \left(\frac{g_\text{bol}}{0.003}\right)\left(\frac{A_{2,10}(\alpha_X)}{0.53}\right)
\left(\frac{\rho_{\bullet\text{a}}}{10^4 \text{M}_\odot h^2 \text{Mpc}^{-3}}\right) \left(\frac{E}{\text{keV}}\right)^{-\alpha_X}\nonumber \\ 
\times & \exp\left[-\left(\sigma_{\text{HI}}(E) + \chi \sigma_{\text{HeI}}(E) \right)N_{\text{H}\bullet}\right]\nonumber \\
\times &\exp\left(-E/300\text{ keV}\right)\text{keV}\,\text{s}^{-1}\,\text{keV}^{-1} h^3\,\text{Mpc}^{-3},
\end{align}

where $N_{\text{H}\bullet}$ is the Hydrogen column depth to the black-holes. In \citet{EwallWice:2018}, we argued that the column depth necessary to explain an EDGES-like absorption feature with a duration of 100\,Myr was roughly $N_{\text{H}\bullet} \approx 10^{23.5}$cm$^{-2}$. In addition, simulations of direct collapse seeds often involve black holes growing in compton thick environments with $N_{\text{H}\bullet}\gtrsim 10^{24}$cm$^{-2}$ \citep{Yue:2013,Pacucci:2015} though it may be difficult for such column densities to be achieved for Pop-III seeds \citep{Alvarez:2009,Smith:2018}.  We adopt $N_{\text{H}\bullet} = 10^{23.5}$cm$^{-2}$ as a fiducial value but we will also explore much lower Hydrogen column depths.  

\begin{table}
\begin{tabular}{|c|l|c|}
Parameter & Description & Fiducial Value \\ \hline
$g_\text{bol}$ & Accreted Mass emitted at 2-10\,keV & $3\times 10^{-3}$\\
$\alpha_{X\bullet}$ & Spectral Index of X-ray emission & 0.9 \\
$N_{\text{H}\bullet}$ & Hydrogen column depth & $3\times 10^{23}$ cm$^{-2}$
\end{tabular}
\caption{Parameters associated with X-ray emission and their fiducial values.}\label{tab:Xrays}
\end{table}

\subsubsection{UV and Optical Emission}\label{sssec:Optical}

UV emission affects the H\,\textsc{i} signal through Ly~$\alpha$ coupling and photo-ionizations. In accreting black-holes, UV photons are expected to arise from thermal disk-emission which is strongly correlated (through IC scattering) with coronal X-rays and usually follows a double power law with a transition at the Lyman limit. Additional emission arises through the reprocessing of absorbed UV and X-ray photons by obscuring gas. Our model includes both reprocessed and primary emission. 

For primary emission from the accretion disk, we adopt the double power-law observed by \citet{Lusso:2015} with $\alpha_{O1} = -0.61$ for $E \leq 13.6$\,eV and  $\alpha_{O2}$ for $200\,\text{eV} > E >13.6$\,eV. We set the overall UV amplitude through its correlation with X-ray emission (in radio-quiet AGN), $\epsilon_{\text{X}\bullet}(2\text{keV})=\epsilon_{\text{UV}\bullet}(2500\mbox{\AA}) 10^{-\alpha_{OX}/0.384}$ and solve for $\alpha_{O2}$ that is consistent with our choice of $\alpha_{OX}$. Combining the values observed in \citet{Lusso:2010} and \citet{Marchese:2012} with the unabsorbed $\epsilon_{\text{X}\bullet}(2\text{keV})$ predicted by equation~\ref{eq:XrayEmissivity} gives us our expression for the UV emissivity of the black-holes, 
\begin{align}\label{eq:UVEmissivity}
\epsilon_{\text{UV}\bullet} (E) & \approx 7.8 \times10^{52} \left(\frac{45\,\text{Myr}}{\tau_\text{s}}\right) \left(\frac{g_\text{bol}}{0.003}\right)\left(\frac{A_{2,10}(\alpha_X)}{0.53}\right) \nonumber \\
& \times  \left(\frac{\rho_{\bullet\text{a}}}{10^4 h^2\text{M}_\odot\text{Mpc}^{-3}}\right) \nonumber \\
& \times 2^{0.9-\alpha_X} \left( 2.48\times 10^{-3} \right)^{1.6-\alpha_{OX}} \left(2.74 \right)^{0.61-\alpha_{O1}} \nonumber \\
&\times \exp\left[ -\left(\sigma_\text{HI}(E) + \chi \sigma_\text{HeI}(E) \right)N_{\text{H}\bullet}\right]\nonumber \\
&  \times\begin{cases} \left(\frac{E}{13.6\text{eV}}\right)^{-\alpha_{O2}}  \times & E>13.6\,\text{eV} \\ \left(\frac{E}{13.6\text{eV}}\right)^{-\alpha_{O1}} & E<13.6\,\text{eV} \end{cases}\nonumber\\
& \times \text{eV}\,\text{s}^{-1}\,\text{eV}^{-1}\,h^3\text{Mpc}^{-3}.
\end{align}
 We list the parameters describing primary UV/Optical emission for our model in Table \ref{tab:Optical} along with their fiducial values.
\begin{table}
\begin{tabular}{|c|l|c|}
 Parameter & Description & Fiducial Value \\ \hline
$\alpha_{OX}$ & Optical-X-ray spectral slope & $1.6$\\
$\alpha_{O1}$ & UV spectral slope & $0.61$
\end{tabular}
\caption{Parameters associated with Optical emission and their fiducial values.}\label{tab:Optical}
\end{table}

The equivalent ionizing escape-fraction for black-hole, with column depth $N_{\text{H}\bullet}$, is  $f_{\text{esc}\bullet} = \int dE \epsilon_{\text{UV}\bullet} / \int dE \epsilon_{\text{UV}\bullet}\exp\left[\left( \sigma_\text{HI}(E) + \chi \sigma
_\text{HeI}(E) \right) N_{\text{H}\bullet} \right]$ which is negligibly small for our fiducial column depths.

For secondary emission, we follow the prescriptions in \citet{Yue:2013} and \citet{Fernandez:2006} in which considers Free-Free/Bound-Free continuum emission and two-photon emission from atoms excited by collisions and recombinations. We solve for the temperature of the obscuring gas following \citet{Yue:2013}, setting $f_e=0.5$ and balancing absorbed power from the black-holes primary spectrum with reprocessed power. 

\subsubsection{Radio}\label{sssec:Radio}
Radio emission from accreting black holes is observed from different phenomena associated with a relativistic jet. These sources include direct emission from the jet itself, which can be highly beamed, emission from $1-10$ pc scale hot-spots where the jet first encounters the intergalactic or intra-cluster medium, and diffuse emission from lobes that fill in a cavity carved out by the jet \citep{Scheuer:1982}. At $\sim$GHz frequencies, the spectral index of core emission is typically flat ($\alpha_R \approx 0$) while hot-spot and lobe emission relatively steep, $(\alpha_R \approx 0.5 - 0.8)$ \citep{Jarvis:2000}. In Faranhoff-Riley type II (FRII) sources, hot-spot emission typically contributes to $\sim 10-100\%$ of the total radio emission \citep{Jenkins:1977}. 

At low redshifts, SE dominates IC scattering within the majority of the radio source since magnetic fields in the hot spots are on the order of $0.1-10$\,mG \citep{Carilli:1991, Fanti:1995} while magnetic fields in the lobes are on the order of $\sim 1-10  \mu$G, with energy densities several hundred times higher then the CMB at $z\lesssim 1$. The lifetime of SE in hot-spots is often $10^3-10^5$ years, requiring continuous injection of fresh electrons by the jet \citep{Carilli:1991} while the lifetimes of SE in radio lobes tends to be on the order of $\sim$\,Myr. A substantial population of Compact Steep Spectrum (CSS) and Gigaherz Peaked Spectrum (GPS) sources also exists at low redshifts with equipartition magnetic field strengths that are often in the 1-10\,mG range \citep{Fanti:1995, Murgia:1999, Murgia:2003}.  


Total radio emission from AGN is often quantified relative to optical emission by a black-hole's ``radio loudness'', $R$, which often refers to the logarithm of the ratio between monochromatic 5 GHz and 2500\,\AA ~(primary) luminosities \citep{Kellerman:1989}. We adopt the definition from \citet{Ivezic:2002} where radio loudness is the log of the ratio between the observed 1.4\,GHz and 8000\,\AA  ~luminosities, translated to the mean redshift of the SDSS-FIRST sample $(z \approx 1)$. Since the mean redshift of the sample is $z \approx 1$. The resulting definition of $R$ is the log-ratio between rest-frame $2.8$\,GHz and 4000\,\AA luminosities.

The nature and evolution of radio-loudness is not yet well understood. Some argue for two underlying populations of ``radio-quiet" AGN (with $R \lesssim 0$) and ``radio loud'' AGN (with $R \sim 3$) \citep{Ivezic:2002,Ivezic:2004}. Others contest that observations of bi-modality arise from selection effects \citep{Cirasuolo:2003,Singal:2011}. We refer to the fraction of sources that are radio loud as $f_L$ and $10^R$ as $\mathcal{R}$
. $\mathcal{R}$ and $f_L$ might evolve significantly over time and there are tentative competing claims of constant, \citep{Banados:2015}, increasing \citep{Donoso:2009,Singal:2011}, and decreasing \citep{Jiang:2007} trends for $f_L$ with redshift. 

If the majority of radio loudness arises from diffuse lobes, then we might expect $\mathcal{R}$ to be lower during the Cosmic Dawn due to IC losses off of the brighter CMB \citep{Ghisellini:2014}. As pointed out by S18, sources at $z \approx 17$ require $\gtrsim$mG magnetic fields to produce appreciable SE. This is not an unusual ask for GPS/CSS sources. However, the majority of CSS/GPS sources are expected to last only $\sim 10^5$ years before their expansion dilutes their magnetic fields below $\sim 1$\,mG \citep{Bicknell:1997, Kaiser:1997}. Hence the sources that we are hypothesizing at CD redshifts must somehow maintain $\gtrsim$\,mG fields over $\gtrsim$\,Myr timescales. We propose several potential ways of doing this in \S~\ref{ssec:IC} but we caution the reader that sustained radio loudness at the level of $z \approx 1$ sources is a generous assumption for sources resembling the radio AGN we observe today. 


The overall goal of this paper is to evaluate how 21\,cm observations might constrain radio loudness at levels similar to $z\approx1$ when X-ray heating and Ly-$\alpha$ coupling from the same sources are realistically modeled self consistently and provide a modeling framework for doing so. Thus, we adopt the optimistic assumption in our fiducial models that radio-loudness is similar to what is observed at the epoch of peak radio-AGN activity. To this end, we adopt the bi-modal radio-loudness distribution in \citet{Ivezic:2002} between {\it rest-frame} 2.78\,GHz and 4000\,\AA\, luminosities. For radio-loud sources, $R$ is distributed as a Gaussian with a mean of $\mu_R \approx 2.8$ and standard deviation $\sigma_R \approx 1.1$\footnote{This is the distribution observed in \citep{Ivezic:2002} for rest-frame radio  $\approx 2.8$\,GHz and optical $\lambda \approx 4000$\,\AA. }. 

Since radio-quiet AGN are typically $\lesssim 10^3$ times fainter then their radio loud counterparts, we do not explicitly include them in our model. To quantify the radio-emission from our black-holes we introduce the parameter 

\begin{equation}
g_\text{R} \equiv \frac{\epsilon_\bullet(2.8 \text{GHz})}{\epsilon_\bullet(4000 \text{\AA})} = 3200 \times \left(\frac{f_L}{0.2}\right)\left(\frac{\langle \mathcal{R} \rangle}{1.6 \times 10^4}\right)
\end{equation}

 Adopting the \citet{Ivezic:2002} distribution radio loudness yields $\langle \mathcal{R}\rangle \approx 1.6\times10^4$. We write down and equation for $\epsilon_{\text{R}\bullet}$ by determining $\epsilon_{\text{UV}\bullet}(\text{4000\AA})$ from equation~\ref{eq:UVEmissivity}, multiplying by $\langle \mathcal{R} \rangle$, and scaling to 1\,GHz. We choose a fiducial value for $f_L$ of 0.2, the local radio loud fraction observed in the most luminous $z\approx 0$ AGN.
 
 Since radio emission at high redshift would likely originate from regions more similar to hot-spots, it is tempting choose their spectral typical index of radio emission, $\alpha_R \approx 0.5$. S18 claim that a $z \approx 17$ radio sources should have $\alpha_R \approx 0.5$. We disagree that this is necessarily be the case (see \S~\ref{ssec:IC}). Thus, we take $\alpha_R = 1.1$ to be our fiducial value but also consider scenarios with $\alpha_R = 0.5$.
  

\begin{align}
\epsilon_{\text{R}\bullet}(\nu) \approx& 7.6\times 10^{23} \left(\frac{g_R}{3200}\right)\left(\frac{g_\text{bol}}{0.003}\right)\left(\frac{A_{2,10}(\alpha_X)}{0.53}\right)\nonumber\\
& \times 2^{0.9-\alpha_X} \left( 2.48\times 10^{-3} \right)^{1.6-\alpha_{OX}} \left(\frac{45\,\text{Myr}}{\tau_\text{s}}\right)\nonumber \\
& \times \left(2.8\right)^{(\alpha_R-0.6)}\left(4.39\right)^{-(\alpha_{O1}-0.61)}\left(\frac{\rho_{\bullet,a}}{10^4 h^2\text{M}_\odot\text{Mpc}^{-3}}\right)\nonumber\\
&\times \left(\frac{\nu}{\text{GHz}}\right)^{-\alpha_R} \text{W}\,\text{Hz}^{-1}\,h^3\,\text{Mpc}^{-3}
\end{align}

\begin{table}
\begin{tabular}{|c|l|c|}
Parameter & Description & Fiducial Value   \\ \hline
$\alpha_R$ & Spectral Slope & $1.1$\\
$g_R$ & Radio gain & 3200
\end{tabular}
\caption{Parameters associated with Radio emission and their fiducial values.}\label{tab:Radio}
\end{table}

\subsection{Illustrative Scenarios}\label{ssec:Models}
\begin{table*}
    \centering
    \begin{tabular}{l|c|c|c|c|c|c|c|c|c|c}
     & Small  & Large  & Large & Un- & Massive  & Highly  & Low-z  & Stars & EDGES  & EDGES   \\ 
    Scenario & Halos & Halos & Fast & Obscured & Seeds & Obscured & Accretion & Only & Small Halos & Large Halos \\ \hline
    $\tau_\text{s}$ (Myr) & 45 & 45 & 20 & 45 & 45 & 45 & $10^3$ & - & 25 & 18 \\
    $\tau_\text{L}$ (Myr)  & $100 $ & $100 $ & $100 $ & $100 $ & $100 $ & $100 $& $100 $& - & $100 $ & $100 $ \\
    $z_\text{min}^i$ & 16 & 16 & 16 & 16 & 16 & 16 & 16& - & 21 & 21 \\
    $T_{\text{vir}\bullet}^\text{min}$ (K) & 2000 & $10^4$ & $10^4$ & $10^4$ & $10^4$ & $10^4$ & $10^4$ & - & 2000 & $10^4$ \\
    $T_{\text{vir}\bullet}^\text{max}$ (K) &  $10^4$ & $5 \times 10^4$ & $5 \times 10^4$ & $5 \times 10^4$ & $5 \times 10^4$ & $5 \times 10^4$ & $5 \times 10^4$  & - & $ 10^4$ & $5 \times 10^4$ \\
    $m_\bullet^i$ (M$_\odot$) & 100 & 100 & 100 & 100 & 1000 & 37 & 100 &  - & 100 & 1500 \\
    $f_\bullet$ & 0.1 & 0.1 & 0.1 & 0.1 & 0.1 & 0.1 & 0.1 & - & 0.1 & 0.1 \\
    $g_\text{bol}$ & 0.003 & 0.003 & 0.003 & 0.003 & 0.003 & 0.003 & 0.003 & - & 0.001 & 0.001  \\
    $\alpha_{\text{X}\bullet}$ & 0.9 & 0.9 & 0.9 & 0.9 & 0.9 & 0.9 & 0.9 & - & 0.5 & 0.5 \\
    $N_{\text{H}\bullet}$ (cm$^{-2}$) & $3 \times 10^{23}$ & $3 \times 10^{23}$ & $3 \times 10^{23}$ & $10^{18}$ & $3 \times 10^{23}$ & $3 \times 10^{24}$ & $3 \times 10^{23}$ & - & $1.8 \times 10^{24}$ & $3 \times 10^{24}$ \\ 
    $\alpha_{OX}$ & 1.6 & 1.6 & 1.6 & 1.6 & 1.6 & 1.6 & 1.6 & - & 1.6 & 1.6 \\
    $\alpha_{O1}$ & 0.61 & 0.61 & 0.61 & 0.61 & 0.61 & 0.61 & 0.61 & - & 0.61 & 0.61 \\
    $\alpha_R$ & 1.1 & 1.1 & 1.1 & 1.1 & 1.1 & 1.1 & 1.1 & - & 1.1 & 1.1  \\
    $\mu_\text{R}$ & 2.8 & 2.8 & 2.8 & 2.8 & 2.8 & 2.8 & 2.8 & - & 2.1 & 2.1 \\
    $\sigma_\text{R}$ & 1.1 & 1.1 & 1.1 & 1.1 & 1.1 & 1.1 & 1.1 & - & 1.1 & 1.1 \\
    $f_\text{L}$ & 0.2 & 0.2 & 0.2 & 0.2 & 0.2 & 0.2 & 0.2 & - & 1 & 1 
    \end{tabular}
    \caption{Values for the Black-Hole growth and emission parameters for all models explored in this work. The parameter choice for ``Small Halos'' corresponds to our ``fiducial'' model.}
    \label{tab:TheOneTable}
\end{table*}
With our semi-analytic framework in place, we can now compute the global 21\,cm signal under the influence of rapidly accreting radio-loud black hole seeds. Listing the parameters in our model (Table~\ref{tab:TheOneTable}), we see that we are adding sixteen. An exhaustive exploration of the phenomenology of every one of these new parameter and their degeneracies is beyond the scope of this work. Instead, we 
 investigate several different scenarios to illustrate the key dependencies of the signal morphology on halo and seed properties, obscuration and growth rate. We do so by focusing on $\tau_\text{s}$, $T^{\text{min/max}}_{\text{vir}\bullet}$, $m_{\bullet}^i$, and $N_{\text{H}\bullet}$ which we vary in the following models. 

\begin{itemize}
\item {\bf Small Halos}:  Adopting all of the fiducial values discussed above, this model corresponds to Pop-III seeds arising in molecular cooling halos that are obscured by dense gas. 
\item {\bf Large Halos}: A Scenario in which black holes form in halos above the atomic cooling threshold. We set $T_{\text{vir}\bullet}^\text{min}=1 \times 10^4$\,K and $T_{\text{vir}\bullet}^\text{max}=5 \times 10^4$\,K. The qualitative impact of larger halo masses is to move the absorption trough to later times. 
\item{\bf Large Halos, Fast}: The same as our large halos scenario except that the accretion time-scale is lowered to $20$\,Myr. Lower $\tau_\text{s}$ primarily increases the depth and side-steepness of the absorption feature while moving it to earlier times. 
\item{\bf Large Halos, Unobscured}: The same as our large halos obscured scenario except that we lower the column densities to $N_{\text{H}\bullet}=10^{18}$\,cm$^{-2}$ which yields and equivalent black-hole UV escape fraction of $0.11$. The primary qualitative effect of reducing $N_{\text{HI}\bullet}$ is to reduce the amplitude of the absorption trough through heating and ionizations.
\item{\bf Large Halos Massive Seeds}: We use this scenario to illustrate potential degeneracies that exist between $m^i_\bullet$ , $\tau_\text{L}$ and $\tau_\text{s}$. In this model, the total energy radiated by a black hole over its lifetime is proportional to $m^f_\bullet = m^i_\bullet e^{\tau_\text{L}/\tau_\text{s}}$ so we can approximately hold the amplitude of the total radio background constant by varying these three parameters in a way that holds $m^f_\bullet$ constant. In our massive seeds model, we start with our fast model and raise $m_\bullet^i$ to $1000\,\text{M}_\odot$ while increasing $\tau_\text{s}$ to keep $m_\bullet^f$ constant. The qualitative effect of increasing $m^i_\bullet$ is similar to decreasing $\tau_\text{s}$. 
\item{\bf Higher Obscuration:} Identical to our ``large halos'' scenario but now with the neutral column depth to black holes raised to $N_{\text{H}\bullet} = 3 \times 10^{24}$\,cm$^{-2}$. 
\item {\bf Low-z Accretion:} A scenario where the accretion properties of the first black holes line up with what is observed in low redshift AGN. Specifically, we set $\lambda=0.1$, $f_\text{duty}=0.1$, and $\epsilon=0.1$ \citep{Shankar:2010,Shankar:2013}. This gives $\tau_\text{s}=4.5$\,Gyr and $g_\text{bol}= 0.006$. We also adopt $\tau_\text{L}=1$\,Gyr \citep{Bird:2008}.  

\item {\bf Stars }: A control scenario with no black holes, $f_\bullet=0$.
\end{itemize}

We show the evolution of $\rho_{\bullet a}$ and $\rho_{\bullet q}$ in Fig.~\ref{fig:Densities} for our various growth scenarios. $\rho_{\bullet a}$ leads $\rho_{\bullet q}$ until $t_\text{max}^i$, when the last black-hole seeds form. At $t_\text{max}^i+\tau_\text{fb}$, the last accreting black-holes shut down, bringing $\rho_{\bullet a}$ to zero and setting $\rho_{\bullet q}$ to a constant value. The onset of density growth is delayed depending on the minimum halo mass while $\tau_\text{s}$ controls its steepness. The much larger accretion times of our low-z models causes them to track the halo-collapse mass early on, which is why our ``low-z'' and ``small halos'' models agree at $t \lesssim 0.125$\,Gyr. All of the models considered produce black-hole densities below the $z\sim 0$ limits of $\lesssim 10^6 h^2$M$_\odot$Mpc$^{-3}$ established by dynamical studies \citep{Merritt:2001}. As we might expect, our ``massive seeds" and ``fast" models produce the same quiescent mass densities, which are set by $m^f_\bullet$ and their identical halo properties (equation~\ref{eq:BlackHoleGrowthQuiescent}). The evolution of accreting densities do differ with the ```massive seeds" scenario assembling  more accretion mass earlier on and with the ``fast" scenario catching up through its faster growth rate.


\section{Modeling Results}\label{sec:ModelingResults}
We now discuss the results of our global signal calculation.

\subsection{The Brightness Temperature and Thermal Evolution}\label{ssec:Temperature}

\begin{figure*}
\includegraphics[width=\textwidth]{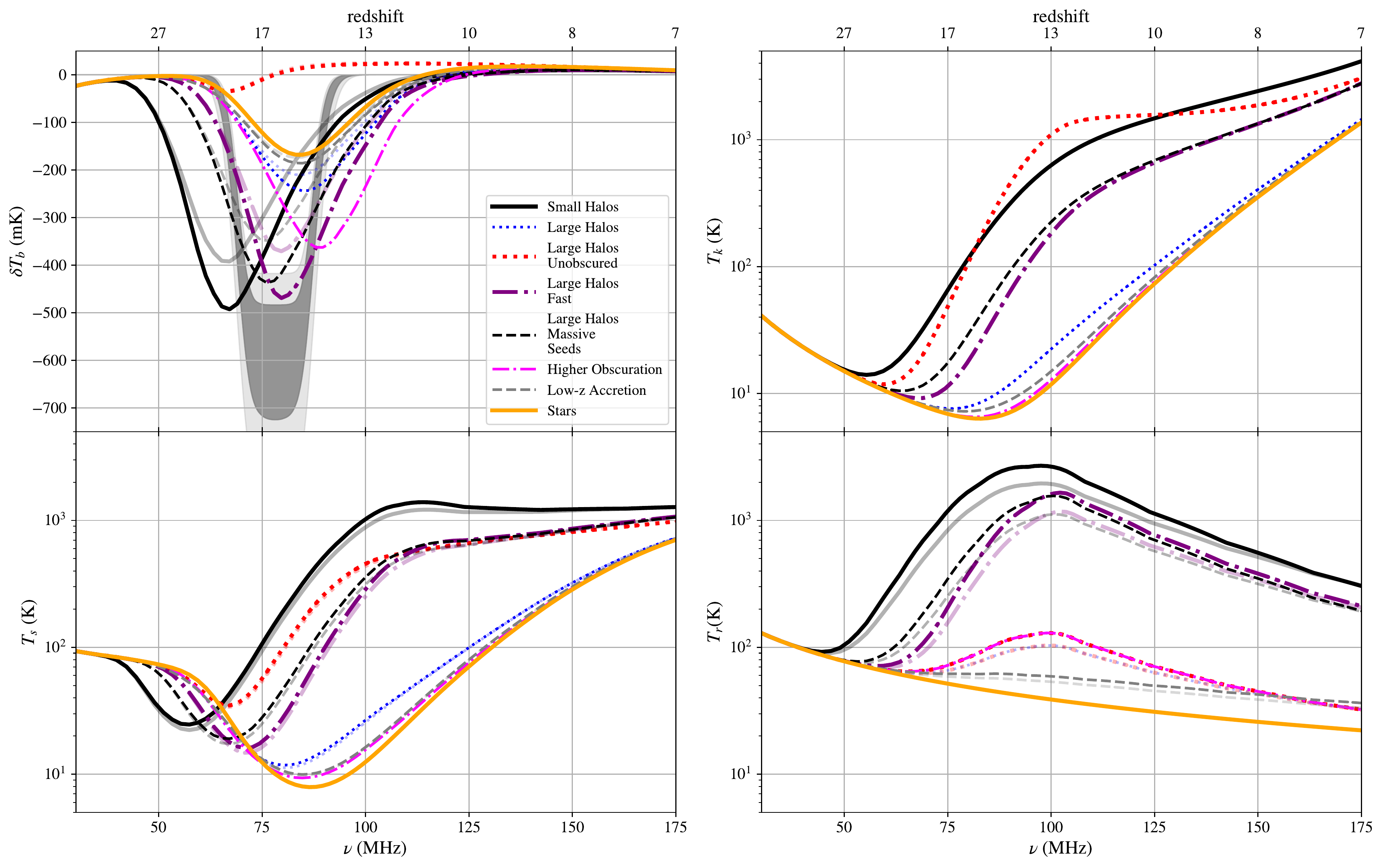}
\caption{The evolution of the 21\,cm brightness temperature, $\delta T_b$ (top left), the radio background temperature, $T_r$ (bottom right), the H~\textsc{i} kinetic temperature, $T_k$ (top right), and the H~\textsc{i} spin temperature, $T_s$, (bottom left) in our simulations. A model without black-holes (orange solid line) exhibits the canonical global signal evolution. Parameters for other models are listed in Table~\ref{tab:TheOneTable}. The grey shaded region shows $1 \sigma$ and $2 \sigma$ contours for the EDGES detection.  Models with black holes introduce a deeper and narrower absorption trough that drops below the $\sim 250$\,mK limit for adiabatically cooling gas absorbing the CMB from recombination.
Increasing the minimum halo mass delays the onset of heating, resulting in an overall translation of the global signal (compare thick black and dotted blue or pink/purple dot dashed lines). Decreasing the Salpeter time, $\tau_\text{s}$, increases the heating and reionization rates, causing the sides of the trough to steepen. Increasing the seed mass both increases the trough depth and translates it to earlier times.
The Small Halos (thin black line) model predicts an absorption feature that is too early to explain EDGES although a 100\,Myr delay between seed formation and accretion can relieve this tension. Light lines denote models where the radio spectral index is flattened from 1.1 to 0.5. 
}\label{fig:Temperatures}
\end{figure*}

We start our discussion with plots of $\delta T_b$ in the top left panel of Fig.~\ref{fig:Temperatures} and compare its timing with the evolution of $T_r$, $T_k$, and $T_s$ in the other panels. 
We show the evolution of $\delta T_b$, under the influence of stars only, as a solid orange line in the top left panel of Fig.~\ref{fig:Temperatures}. Without black holes, $\delta T_b$ follows the canonical evolution observed in most theoretical models with an absorption trough initiated by Ly~$\alpha$ coupling of the spin temperature to the kinetic temperature of the adiabatically cooled H\,\textsc{i} gas and ended by X-ray heating which drives the H\,\textsc{i} into emission before ionization brings $\delta T_b$ to zero. 

All of our models that involve obscured black-holes yield a feature that is significantly deeper and narrower than our stellar model (Fig.~\ref{fig:Temperatures}, top left, excluding the black dotted, blue dashed, and orange solid lines). Many of the scenarios, which either involve greater obscuration or enhanced black hole densities, extend beyond the $\sim 250$\,mK limit that is expected without a radio background (top left purple/pink dot-dashed and solid black lines). 

Comparing scenarios with large (all lines but orange solid) and small (thick black line) halos, one sees that raising the minimum halo virial temperature causes an overall delay in the evolution of $\delta T_b$. We see that the Small Halos model, which roughly corresponds to SMBH progenitors forming from Pop-III seeds, predicts an absorption trough that is too early to explain the EDGES observation. This is consistent with \citet{Kaurov:2018}'s conclusion that the timing of the EDGES trough suggests $Ly~\alpha$ coupling and heating driven by massive halos. However, a 100\,Myr delay between Pop-III seed formation and accretion, which has been predicted by some models \citep{Johnson:2007}, can shift the feature such that the timing of the absorption feature agrees with EDGES.

The spectral index of $\alpha_R = 1.1$ depends on whether electrons stay indefinitely in synchrotron emitting regions or advect into regions where they lose a significant fraction of their energy through IC losses (see \S~\ref{ssec:IC}). Thus, we also look into the consequences of a flatter spectral index by plotting models with $\alpha_R=0.5$ as light lines in Fig.~\ref{fig:Temperatures}. By reducing the rest-frame emission below 2.8\,GHz, flatter spectral indices tend to reduce the radio backgrounds and absorption amplitudes in our models.

\begin{figure}
\includegraphics[width=.5\textwidth]{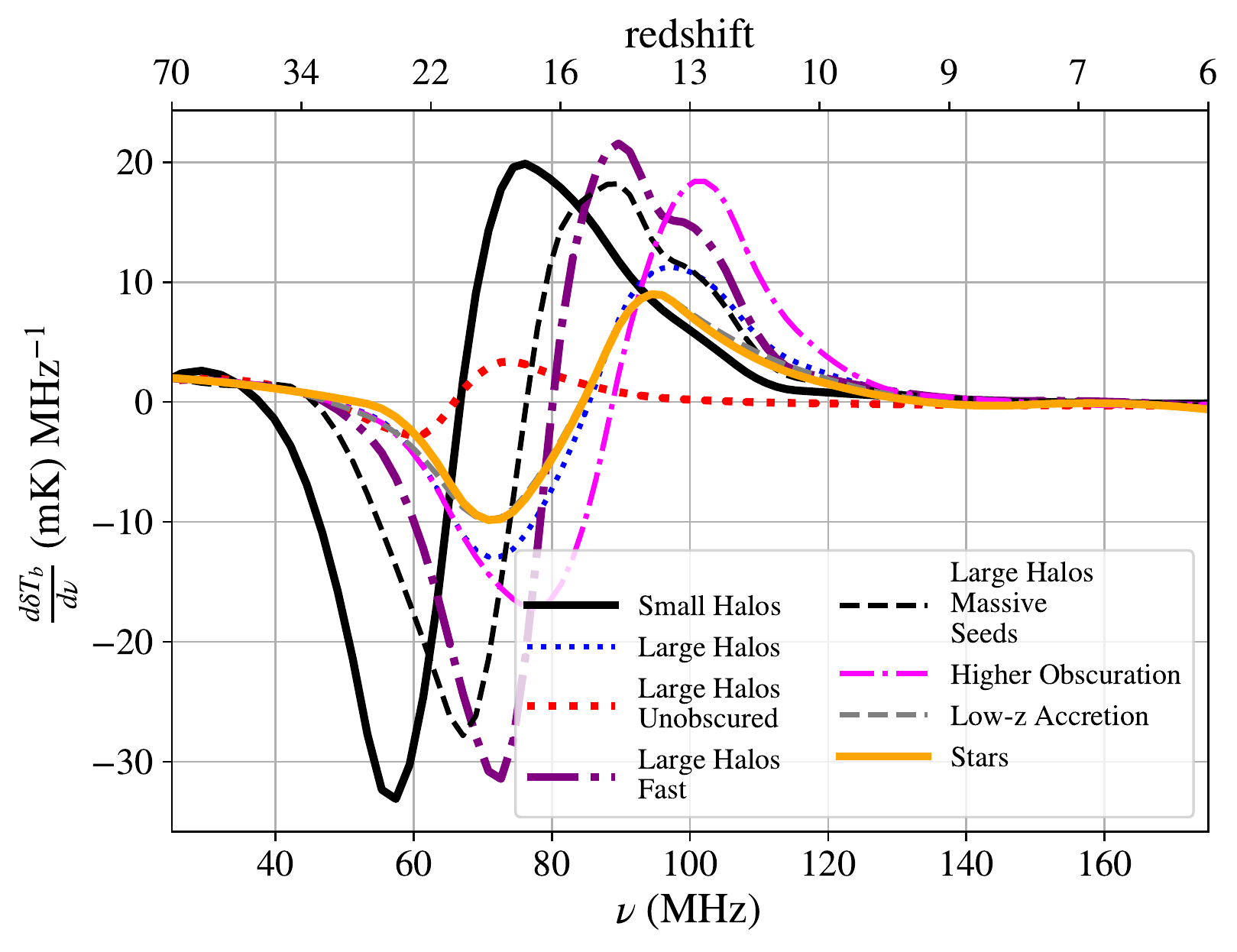}
\caption{The derivative of $\delta T_b$ with respect to frequency. 
Comparing the Large Halos and Fast models we see that the steepness of the rising and falling sides of the sides of the absorption feature are increased with decreasing $\tau_\text{s}$. 
}\label{Fig:dTb_dnu}
\end{figure}

The contrast between our Large Halos (dotted blue) and Large Halos Fast (purple dot dashed) models illustrates how $\tau_\text{s}$ affects the steepness of the absorption feature's sides. Smaller $\tau_\text{s}$ corresponds to a deeper and narrower trough due to faster Ly~$\alpha$ coupling, heating, and radio emission. We illustrate the higher rate in evolution that results by plotting the derivative of $\delta T_b$ with respect to frequency in Fig.~\ref{Fig:dTb_dnu}.

 Inspecting the up-turn at $z\sim 17$ in the evolution of $T_k$ and $T_s$ in the top-right and bottom-left  panels of Fig.~\ref{fig:Temperatures}, we see that the Ly~$\alpha$ emission from black holes and stars only partially couples the spin temperature to $T_k$ at its minimum. Since the $T_s$ curve in Fig.~\ref{fig:Temperatures} does not reach as low as $T_k$ coupling of $T_s$ to $T_k$ does not happen until substantial X-ray heating has already occurred. In many of the more radio emissive scenarios, enhanced rates of absorption and stimulated emission of radio-background 21\,cm photons prevent complete Ly~$\alpha$ coupling from ever occurring. 
 
The brightness temperature evolution for our Massive Seeds (thin dashed black line) and Fast (purple dost-dashed line) models are similar since both radiate equal amounts of energy per co-moving volume. However, the Massive Seeds scenario shifts the total brightness evolution to slightly earlier times since more massive seeds result in more mass being assembled at earlier times than the Fast model, even if both scenarios eventually assemble the same black hole mass. Increasing the obscuration of black holes to reduce and delay heating is another way of increasing the total absorption depth. Comparing the Higher Obscuration (pink dot dashed), Massive Seeds (thin dashed black), and Fast (purple dot-dashed line) curves in Fig.~\ref{fig:Temperatures}, we see that increasing the obscuration depth also moves the trough to lower redshifts (the opposite effect of Massive Seeds). 

Comparing $T_k$ and $\delta T_b$ for all of our models (Fig.~\ref{fig:Temperatures}), we see that the absorption minimum occurs after some X-ray heating has already taken place and $T_r$ is within a factor of a few of the CMB value. V18 find an order $10\%$ change in $\delta T_b$ from radio heating for a radio background that is $3.5 \times$ the CMB value with no X-ray heating. Since the gas in our model experiences similar CMB levels at $z \gtrsim 17$, in addition to X-ray heating, we conclude that the V18 effect has a $\lesssim 10\%$ impact on the amplitudes of our predicted absorption amplitudes for $z \gtrsim 17$. 

All of our black-hole scenarios accelerate the global-signal's evolution beyond the stellar scenario. This is primarily due to the fact that the black-hole emissivities, through exponential growth and large seed masses, can outpace the time-evolution of stellar emissivities which are limited by the halo-collapse rate.

 \subsection{Ionization Histories}\label{ssec:ModelingReionization}

\begin{figure}
\includegraphics[width=.5\textwidth]{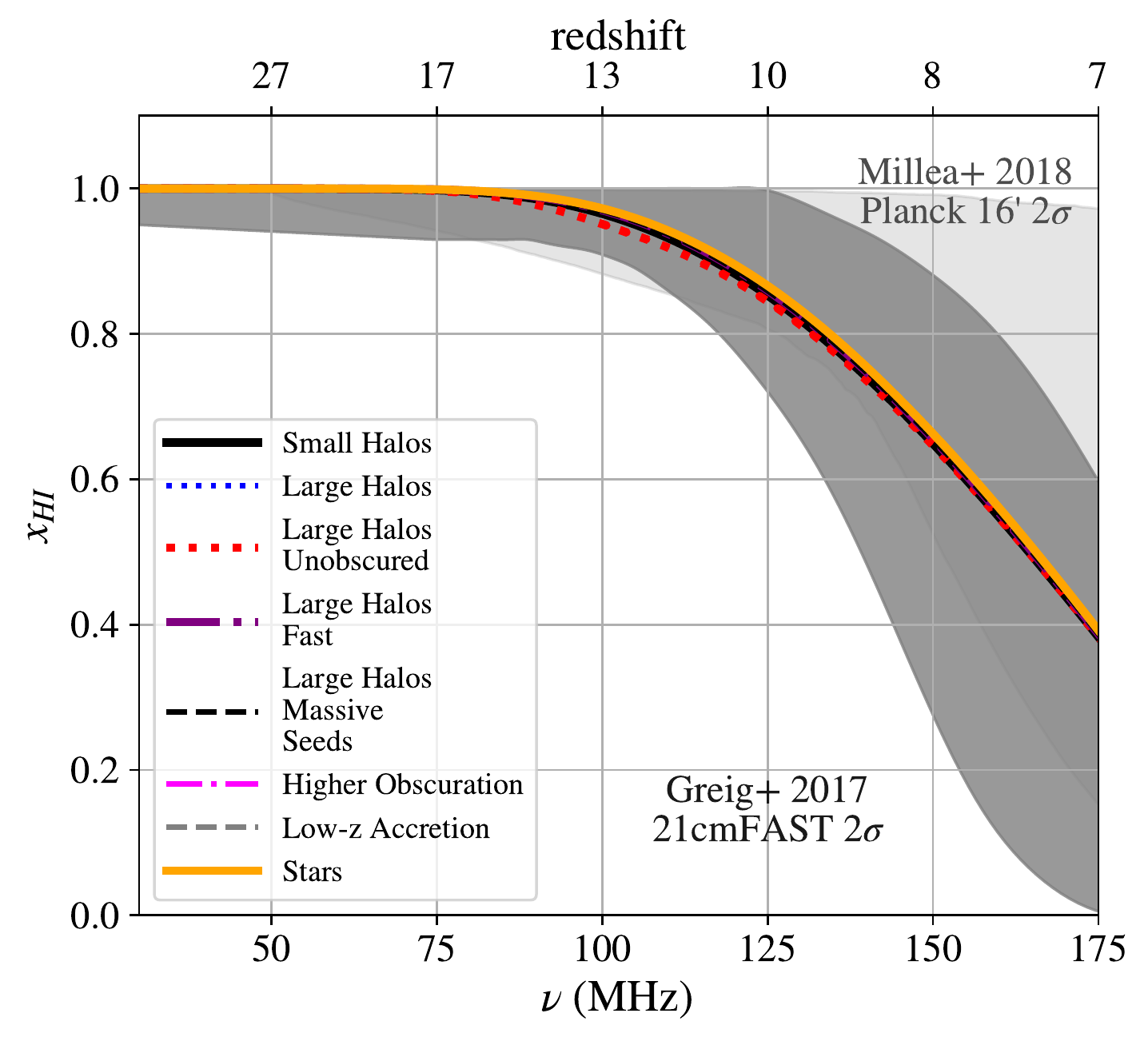}
\caption{The evolution of the neutral fraction, $x_\text{HI}$. We include the {\tt 21cmFAST} derived 2$\sigma$ contours from \citet{Greig:2017} (dark grey contours) along with the PC derived constraints from \citet{Millea:2018} (light grey contours).}\label{fig:XHI}
\end{figure}

Fig.~\ref{fig:XHI} shows the ionization histories of our models. All are similar though the reinization produced by our ``unobscured'' model is slightly more rapid. 
To get a sense of how our models line up against existing constraints, we show the $2\sigma$ region from \citet{Greig:2017} which is derived by fitting a popular three-parameter model to observations of the CMB and quasar specta. We also show the $2\sigma$ contours from the model-independent principal component (PC) analysis of \citet{Planck:2016} data derived by \citet{Millea:2018}. A ll of our models are consistent with the \citet{Millea:2018} and \citet{Greig:2017} constraints.

\subsection{Local Backgrounds}\label{ssec:Backgrounds}

\begin{figure}
\includegraphics[width=.5\textwidth]{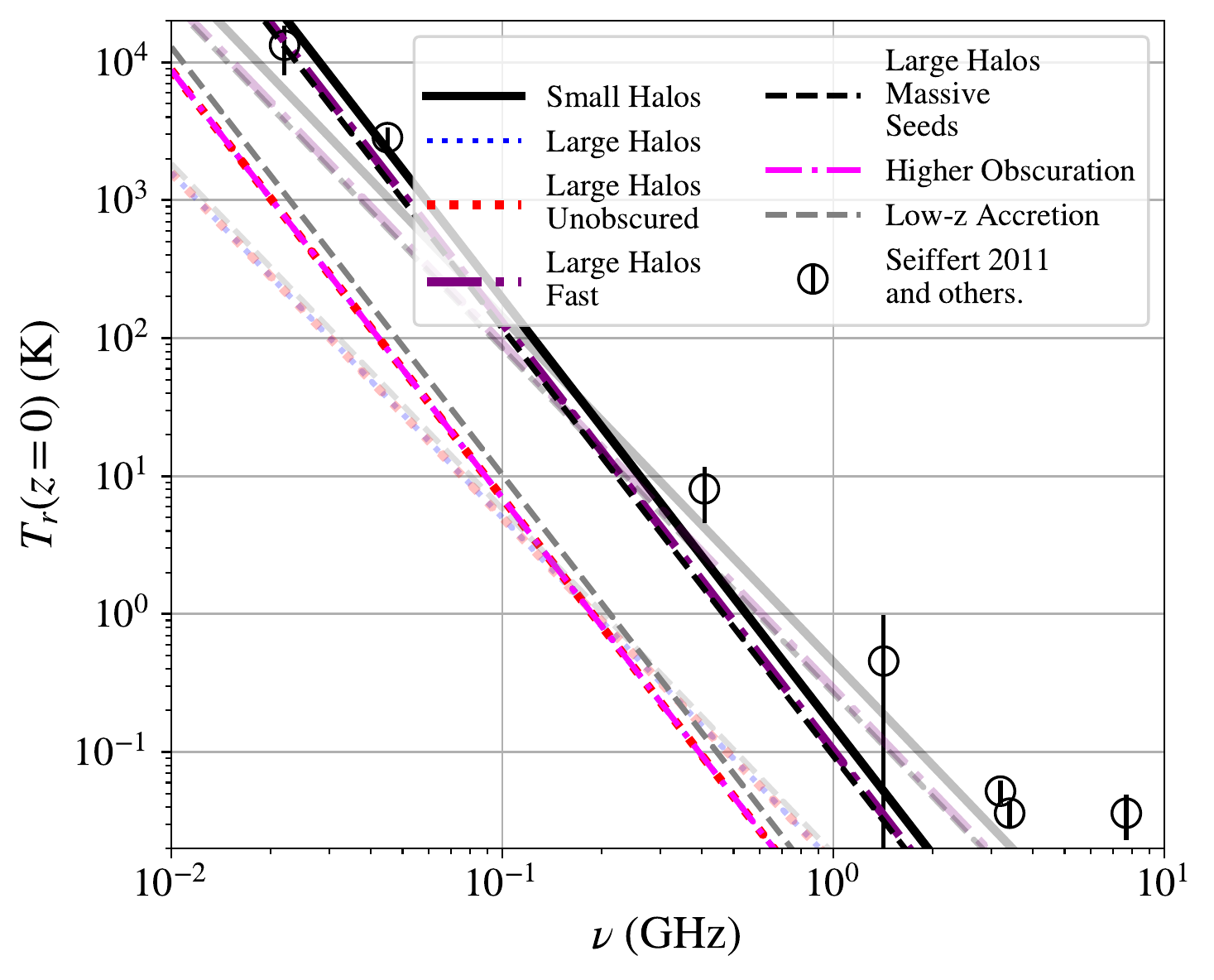}
\caption{The radio-background predicted for our models as a function of frequency compared (various lines) to measurements of the radio monopole with point sources subtracted (black circles) from \citet{Seiffert:2011} and references therein. The models that we consider produce radio-backgrounds that are below or consistent with existing constraints. Light lines denote the same models as the dark lines but with a flatter spectral index ($\alpha_R = 0.5$). We do not include our Stars Only model in this plot since it does not produce any radio emission.}\label{fig:RadioBackground}
\end{figure}

How do cosmological backgrounds vary across our models? In Fig.~\ref{fig:RadioBackground}, we plot the radio monopole observed at $z=0$ as a result of our black hole models and compare them to various measurements (\citet{Seiffert:2011} and references therein). As one might expect, the models that achieve a larger absorption trough by increasing co-moving radio emissivity also result in the largest backgrounds. Since ``massive seeds'' (thin dashed black line) and ``fast'' (purple dot dashed line) models both radiate the same amount of energy in the radio so their aggregate backgrounds are practically identical. This is despite the fact that the ``massive seeds'' model yields a slightly earlier trough (Fig.~\ref{fig:Temperatures}). Since they involve identical black hole masses and primary emission properties, the ``Large Halos'', ``Unobscured'', and ``Higher obscuration'' models also produce identical radio backgrounds and different absorption signatures. We note that at the column depths considered, obscuring gas only has a small impact on radio propagation \citep{EwallWice:2018}.

\begin{figure}
\includegraphics[width=.5\textwidth]{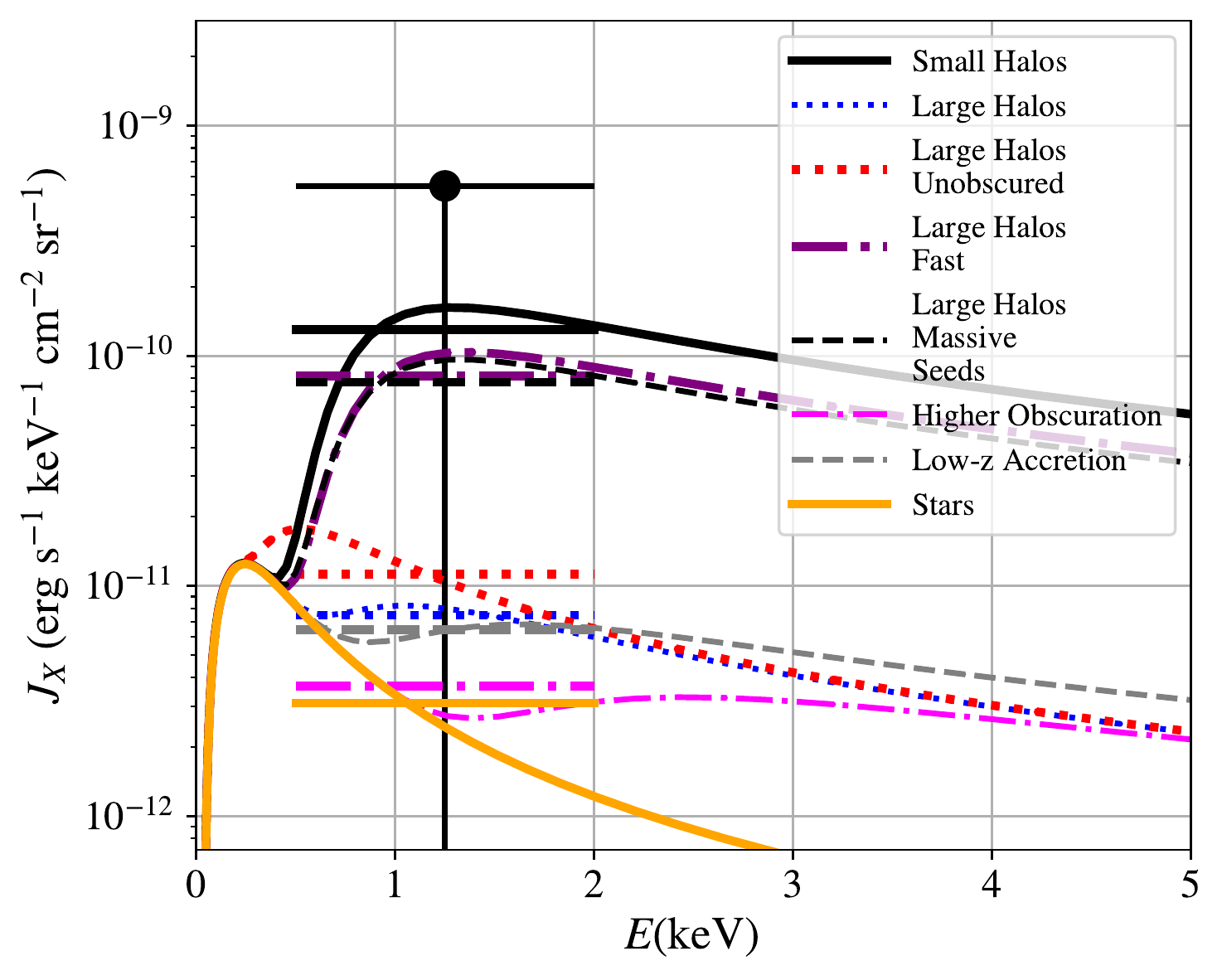}
\caption{The soft XRB arising from our various global signal models described in \S~\ref{ssec:Models}. Obscured models agree well with their unobscured counterparts at high X-ray energies (compare red dotted and blue dotted lines). The black point indicates the upper limit on the unresolved extra-galactic background implied by Chandra deep field limits between 0.5 and 2 keV. Horizontal lines are the average brightness over $0.5-2$\,keV for each model. The large obscuring column depths required for AGN to explain EDGES imprint a distinctive double-peaked Cosmic Dawn XRB where the low energy peak arises from less obscured X-ray binaries. }\label{fig:XRB}
\end{figure}

We also compute the soft X-ray background (XRB) for each model which we plot in Fig.~\ref{fig:XRB}. The disparate obscuration depths for XRBs and AGN yield a distinctive peaked structure at low energies. Since high obscuration of AGN is required to explain the EDGES feature, limits on any double peaked nature of the Cosmic Dawn XRB might help validate or constrain black-hole accretion as an explanation. The amplitude trends noted in the radio backgrounds hold the same for X-rays except for the absence of an obscuration related cutoff since varying $N_{\text{HI}\bullet}$ does not affect the emergent radio spectrum. We compare predicted XRBs to the $\sim 2.5 \times 10^{-13}$\,erg\,s$^{-1}$\,cm$^{-2}$\,deg$^{-2}$ upper limit on the 0.5-2\,keV unresolved extra-galactic XRB determined by \citet{Cappelluti:2013,Fialkov:2017} and find that none of our models exceed this limit.

\subsection{The Impact of Radio Loudness on the Global 21\,cm Signal}\label{ssec:RadioLoudness}

\begin{figure}
\includegraphics[width=.5\textwidth]{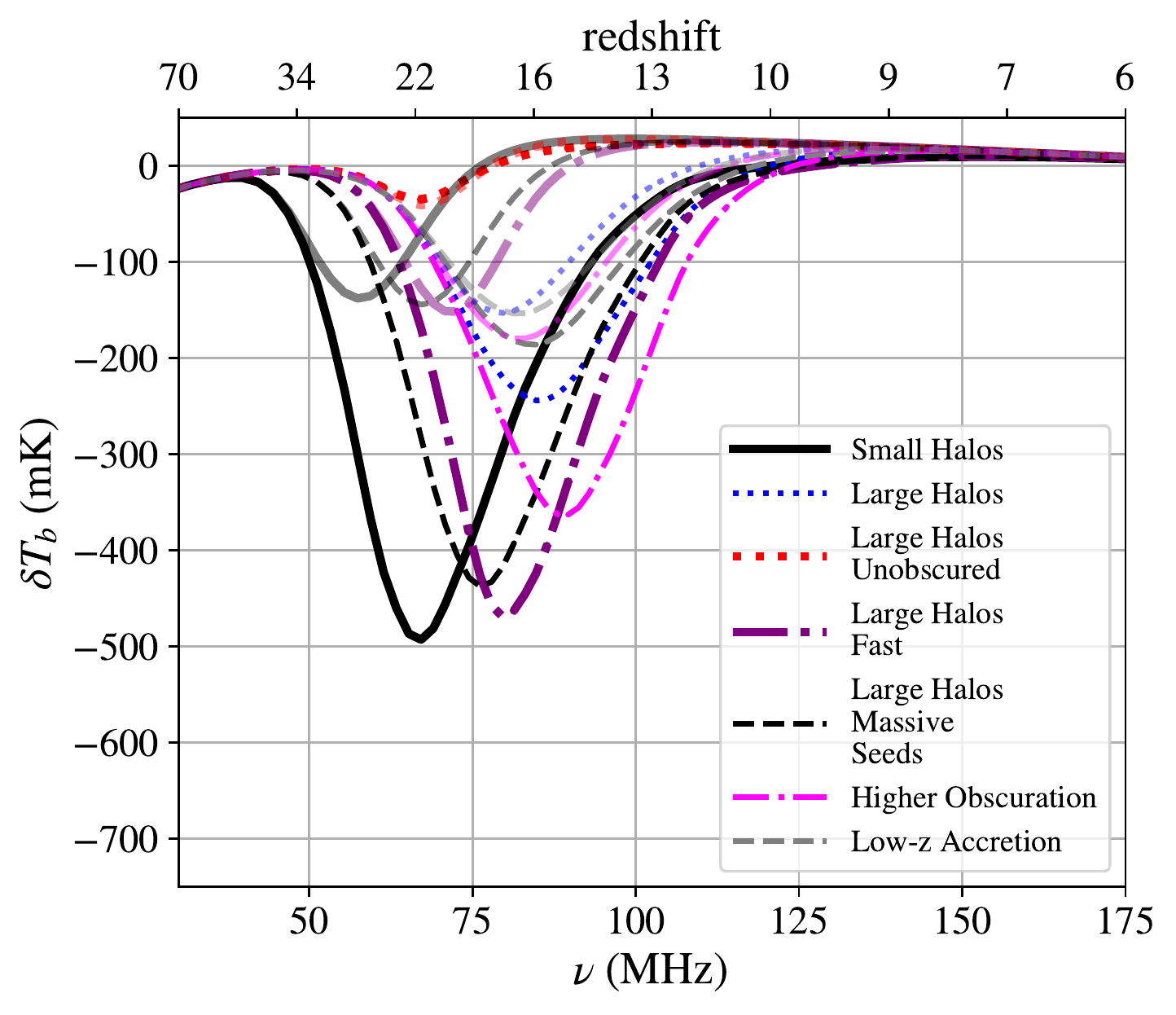}
\caption{The 21\,cm brightness temperature for our various black-hole scenarios with a radio fraction of  $f_L=0.2$ (dark lines) and $f_L=0.0$ (light lines). In scenarios with high accretion rates, the 21\,cm signal arising from obscured black-holes can be impacted at the hundreds of percent level.  Radio-loudness can have a $\sim 10\%$ impact (grey dashed lines) even when black-hole seeds experience accretion rates similar to what is observed at low redshifts. }\label{fig:RadioLoudnessComparison}
\end{figure}

We examine the level that the radio-loudness of AGN impacts $\delta T_b$ as a first attempt to understand at what level 21\,cm global-signal measurements might constrain the existence of radio loud accretion during the Cosmic Dawn. In Fig.~\ref{fig:RadioLoudnessComparison}, we compare $\delta T_b$ for our black-hole scenarios with $f_L=0$ and $f_L=0.2$. While radio-emission has little impact on scenarios dominated by unobscured black-holes, its impact on obscured models ranges from tens to hundreds of percent; even when the accretion rates and duty cycles are at relatively low levels. It is therefore important to include the impact of radio emission when the 21\,cm signal is heavily impacted by the growth of black hole seeds, such as those considered by  \citet{Zaroubi:2007} and \citet{Tanaka:2016}. When a fraction of AGN $(\sim 10\%)$ are radio loud, the trough appears deeper and later than it would otherwise (compare dark and light sets of lines). It is likely that lower levels of accretion will be degenerate with other astrophysical signatures but should be included to account for such degeneracies.

 While an absorption trough that is deeper than the adiabatic minimum would be suggestive of excess radio emission, it is unclear whether the differences caused by radio emission can be disentangled from other astrophysical effects when this is not the case. That the power-spectrum rises to particularly large amplitudes at large $k$ in the presence of radio-loud black-holes may be used to break this degeneracy \citep{Ewall-Wice:2014}. 

\subsection{Source Counts}\label{ssec:Source Counts}
\begin{figure*}
\includegraphics[width=\textwidth]{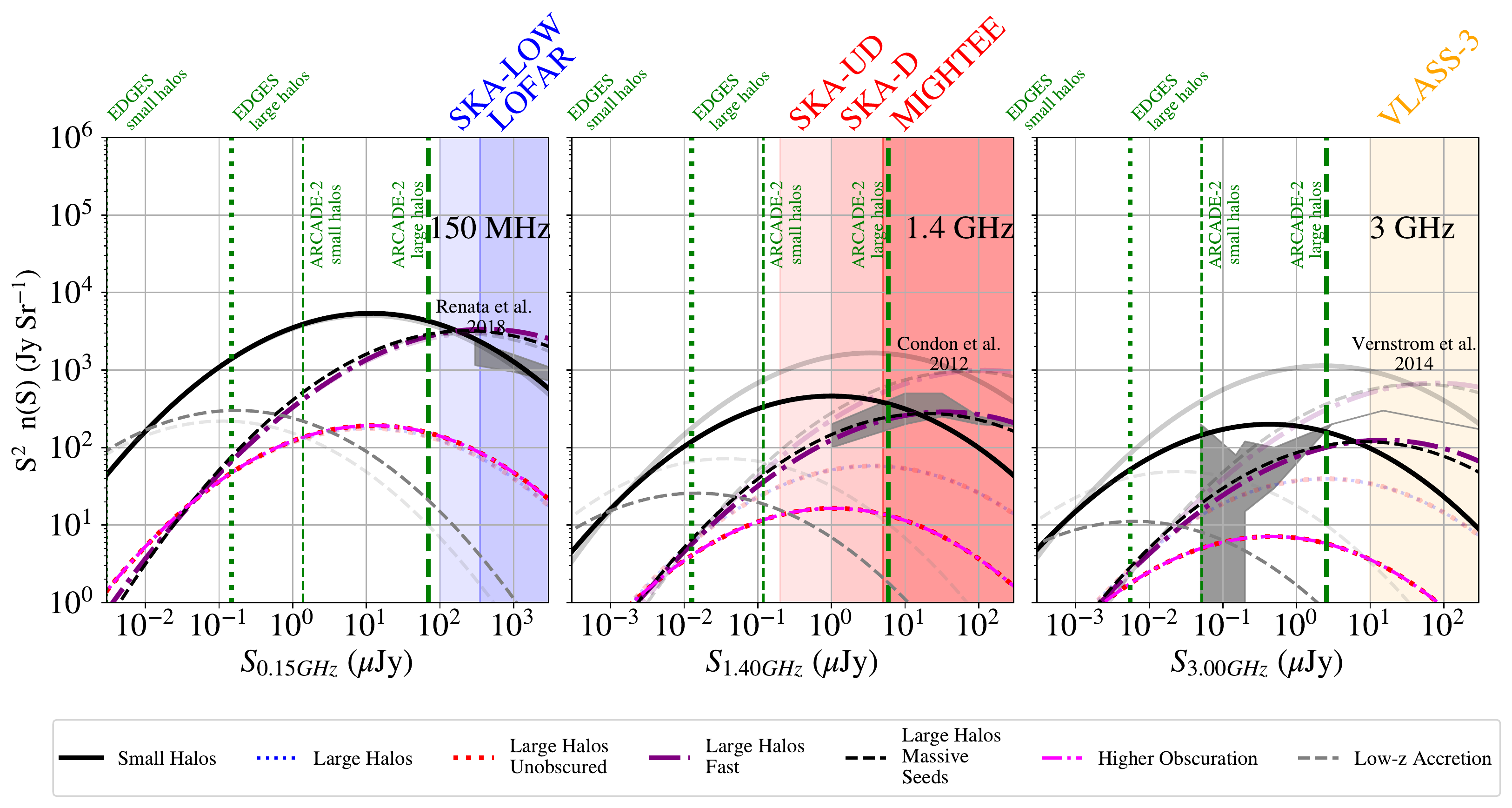}
\caption{Intensity per logarithmic flux-bin for our various black-hole models. The grey shaded regions denote constraints derived from the confusion and number count analyses by R18 at 150\,MHz, C12 at 1.4\,GHz and V14 at 3\,GHz which exclude the Small Halos Fast model due to its large number of predicted sources. We note that this would not be the case for a tighter radio loudness distribution (Fig.~\ref{fig:dndsDelta}). Since the Fast and Massive Seeds models produce most of the observed 1.4\,GHz source counts between $1-10$\,$\mu$Jy, they may be in tension and/or are degenerate with contributions from SFGs.  While the Fast, and Massive Seeds scenarios predict similar radio backgrounds, their flux distributions are quite different. Thus, radio surveys are helpful in removing degeneracies that exist in observations of $\delta T_b$ alone. Fluctuation analyses and expected SFG counts already constrain the Small Halos, Fast, and Massive Seeds models. We do not include the Stars  model in this plot since it does not produce any radio point sources. The projected 5$\sigma$ point-source detection thresholds of point-source surveys by the SKA1, LOFAR, and the VLA calculated in P15 are denoted by colored shaded regions. Green vertical lines denote the characteristic fluxes of sources that can explain the ARCADE-2 (vertical dashed green lines) and EDGES (vertical dotted green lines) detections. Transparent sets of lines show models where the spectral index has been flattened from $1.1$ to $0.5$. Flatter spectrum scenarios are generally more constrained then their steep spectrum counterparts.  
}
\label{fig:dnds}
\end{figure*}

We compute radio-source counts to determine whether future surveys might be used to test 
our models and help lift degeneracies in the global signal. In Fig~\ref{fig:dnds}, we show the contribution of sources per decade of flux towards the total background intensity which can be written as $dn/d S S^2$ where $dn/dS$ is the differential number of sources per unit solid angle and unit flux. 

\begin{figure*}
    \centering
    \includegraphics[width=\textwidth]{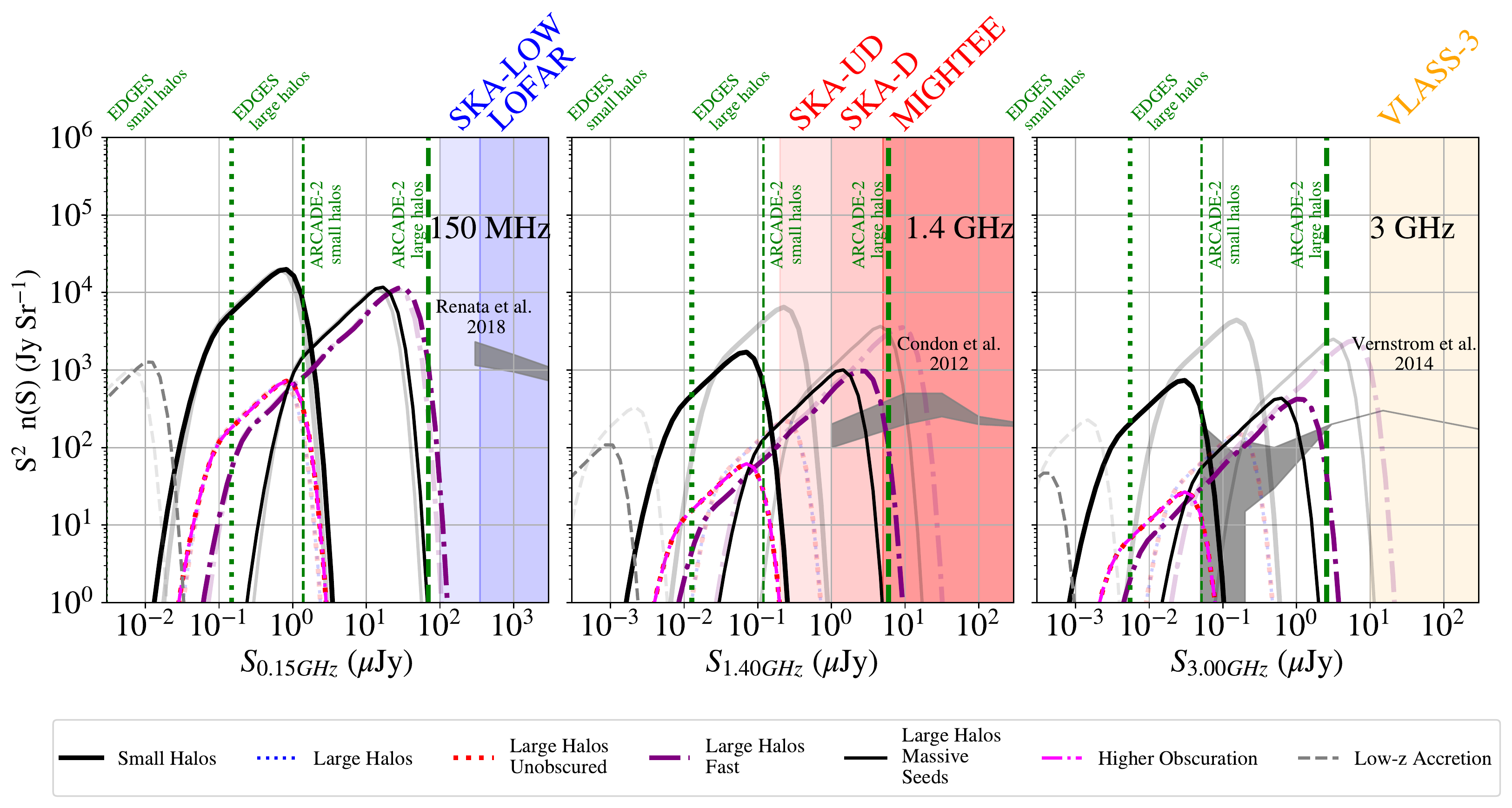}
    \caption{The same as Fig.~\ref{fig:dnds} but now collapsing the log-normal radio loudness distributions into delta functions that yield the same mean co-moving radio emissivity. This removes the high flux tails in Fig.~\ref{fig:dnds} and increases the number of sources at the distribution peaks. This trade-off allows for models involving molecular cooling halos to skirt below (albeit in a contrived way) the R18/C12/V14 constraints. }
    \label{fig:dndsDelta}
\end{figure*}

To understand how much of the scatter in source fluxes arises from the width of the radio loudness distribution, we also compute $S^2 dn/dS $ when the radio-loudness distribution is a delta-function at the loudness that gives the same co-moving emissivity as the usual log-normal (Fig.~\ref{fig:dndsDelta}). Such a delta distribution would be practically indistinguishable from the log-normal in mean background measurements. We find that eliminating the spread in radio-loudness significantly tightens up the radio flux distributions while the flux distribution peaks remain the same.

 We also examine flux distributions for a shallower spectral index of $\alpha_R = 0.5$ (light lines in Figs.~\ref{fig:dnds} and \ref{fig:dndsDelta}). While our steep spectrum is motivated by predictions for synchrotron aging S18, a spectral index of $0.5$ is more consistent with the ARCADE-2 excess. A shallower spectral index results in higher source counts at $\gtrsim$\,GHz frequencies. 
 
 While the ``Large Halos Fast'', ``Massive Seeds'', and ``Small Halos'' models yield similar total intensities (Fig.~\ref{fig:RadioBackground}), they are highly separated in Fig.~\ref{fig:dnds} and Fig.~\ref{fig:dndsDelta}. Specifically, these models have different peak fluxes. This is because the maximum source luminosity is proportional to $ \tau_\text{s}^{-1} m_\bullet^i e^{\tau_\text{s}/\tau_\text{L}}$. This quantity is lowered in the ``Massive Seeds'' model which at the same time has the same $m_\bullet^i e^{\tau_\text{s}/\tau_\text{L}}$ and larger $\tau_\text{s}$. 

\begin{table}
\begin{tabular}{l|l|l|l}
Name & Frequency & 5-$\sigma$ threshold &  Area\\
\hline LOFAR & 150\,MHz & 350 \,$\mu$Jy & All-Sky \\
SKA1-LOW & 150\,MHz & 100\,$\mu$Jy & All-Sky \\
MIGHTEE & 1.4 \,GHz & 5\,$\mu$Jy & $20$\,deg$^2$ \\
SKA1-DEEP & 1.4\,GHz & 1\,$\mu$Jy & 10 deg$^2$ \\
SKA1-ULTRADEEP  & 1.4\,GHz & 0.1\,$\mu$Jy & 1\,deg$^2$ \\
VLASS-3 & 3\,GHz  & 10\,$\mu$Jy & 10\,deg$^2$
\end{tabular}
\caption{The properties of future and in progress radio-surveys including LOFAR \citep{Shimwell:2019}, MIGHTEE \citep{Jarvis:2016}, VLASS-3 and the SKA \citep{Prandoni:2015}}
\label{tab:SurveySensitivities}
\end{table}

A number of our models, while allowed in background measurements, are eliminated by existing limits on point source populations. To illustrate this, we plot the $2-\sigma$ regions allowed by recent fluctuation/number count analyses at $150$\,MHz \citep{Retana:2018} (R18) (also see \citealt{Williams:2016,Hardcastle:2016}), $1.4$\,GHz \citep{Condon:2012} (C12) and $3$\,GHz \citep{Vernstrom:2014} (V14). When wide radio-loud distributions are present, their high flux tails tend to conflict with existing limits if most $\sim 10 \mu$Jy sources are indeed SFGs (and potentially confuse interpretations if these sources are not). For narrow radio loudness distributions, the Fast and Massive Seeds models discussed here are still ruled out while the other scenarios are allowed since they concentrate a large number of sources below the thressholds of current fluctuation analyses.

To determine the detectability of Cosmic dawn black holes in future radio surveys, we also show the flux-sensitivities of future experiments from \citet{Jarvis:2016} and \citet{Prandoni:2015} (P15) in Figs.~\ref{fig:dnds} and \ref{fig:dndsDelta}.
These sensitivities take into account thermal noise and the confusion limit of known power-law distributions of sources. We have found that the contributions of the CD black holes to confusion noise to be negligible in comparison.

Planned deep surveys on the mid band SKA~1 (SKA1-MID) will resolve the flux peaks of models producing an EDGES level feature with sources located in atomic cooling halos, but not necessarily the scenarios with small Pop-III black holes in molecular cooling halos. While it may not be possible to resolve individual sources in these models, fluctuation analyses might still be used to constrain scenarios driven by fainter $\lesssim \mu$\,Jy populations. 

 All of the models that we examine that produce an EDGES level feature also introduce a substantial $\sim \mu$Jy source population. However these sources can, in principal, be moved to smaller (and more numerous) halos in exchange for smaller fluxes.  

The minimum halo mass hosting a black hole can be used to set a characteristic flux for sources that produce a radio background with temperature $T_\text{ref}$ at redshift $z_\text{ref}$ and frequency $\nu_\text{ref}$. For an order-of-magnitude calculation, we assume that all of our radio sources have equal luminosity, occupy $f_\text{halo}$ of dark-matter halos between $T_\text{vir}^\text{min}$ and $T_\text{vir}^\text{max}$, and exist up to redshift $z_\text{min}$. To generate a radio background of $T_\text{ref}$ at $z_\text{ref}$, their distribution must be overved at $z=0$ to include sources with fluxes of at least 
\begin{align}\label{eq:Smin}
    S_{\nu_\text{obs}}^\text{min} &= \frac{2 k_B T_\text{ref}}{\lambda_\text{ref}^2(1+z_\text{ref})^{1+\alpha_R}} \left(\frac{\nu_\text{obs} (1+z_\text{min} ) }{\nu_\text{ref}} \right)^{-\alpha_R}  \nonumber \\
    &\times \left[ c D_L^2(z_\text{min}) \int_{z_\text{min}}^\infty  \frac{n(z^\prime) dz^\prime}{H(z^\prime) (1+z^\prime)^{1+\alpha_R}}\right]^{-1}.
\end{align}
In reality, the luminosities of black holes should increase with time so that the most luminous sources appear at lower redshifts with even larger fluxes. Hence, we can consider the characteristic flux obtained from equation~\ref{eq:Smin} as an order of magnitude lower bound on source fluxes required to produce $T_\text{ref}$.

To produce the EDGES feature, the radio background must at least be as bright as $T_\text{ref} \gtrsim 2.73 \text{K} (1+z_\text{min}) $ at $\nu_\text{ref} \approx 1.4 \text{GHz}$ with $z_\text{min} = z_\text{ref} = 17$. Plugging these numbers into equation~\ref{eq:Smin}, we obtain lower limits on flux densities in Fig.~\ref{fig:dnds} which we denote with vertical dotted green lines for ``large'' and ``small'' halos. We see that producing the EDGES excess with atomic cooling halos requires a population of sources with $S \sim 10^{-2} \mu$\,Jy at GHz frequencies while if molecular cooling halos hosted the same sources, the characteristic fluxes of these sources is on the order of $S \sim \text{nJy}$. From the dashed green lines, source fluxes to produce the ARCADE-2 excess at 3\,GHz require a population of $\sim 10^{-1} - 10^{1}$\,$\mu$Jy sources at $\sim 1.4$\,GHz. Since most of our scenarios produce radio-backgrounds similar to ARCADE-2, they simultaneously give rise too source populations with characteristic fluxes within an order of magnitude of our predictions. While these characteristic fluxes are sensitive to our choice of spectral index (1.1), and the specific fraction of halos hosting black holes (we chose 10\%) they serve as an order of magnitude estimate of what sort of fluxes we should expect from radio surveys if EDGES or ARCADE-2 are produced by discrete sources with one source per halo. 

\section{Discussion}

\subsection{Is Inverse Compton Cooling a Showstopper?}\label{ssec:IC}

In all of our models, the global signal is heavily impacted only when the ratio between radio and bolometric luminosity is similar to low redshift. As recognized in previous works \citep{Ghisellini:2014,Ghisellini:2015,Saxena:2017}, this cannot be the case for classical FRII lobes where magnetic fields tend to be below $100\mu$G. More recently, S18 showed that regions where magnetic fields are below $\lesssim 1000\mu$G would not produce appreciable SE due to IC scattering. We agree with S18 that the sorts of radio sources that significantly impact the 21\,cm signal need to have $\gtrsim$\,mG magnetic field strengths. S18 go a step further however, claiming that even assuming $\gtrsim$\,mG fields, sources at $z\approx17$ would have to be $\sim 1000$ times more radio loud then today to produce similar levels of radio emission. We do not think this second argument to be true as we will now explain. 

\begin{figure}
    \centering
    \includegraphics[width=.5\textwidth]{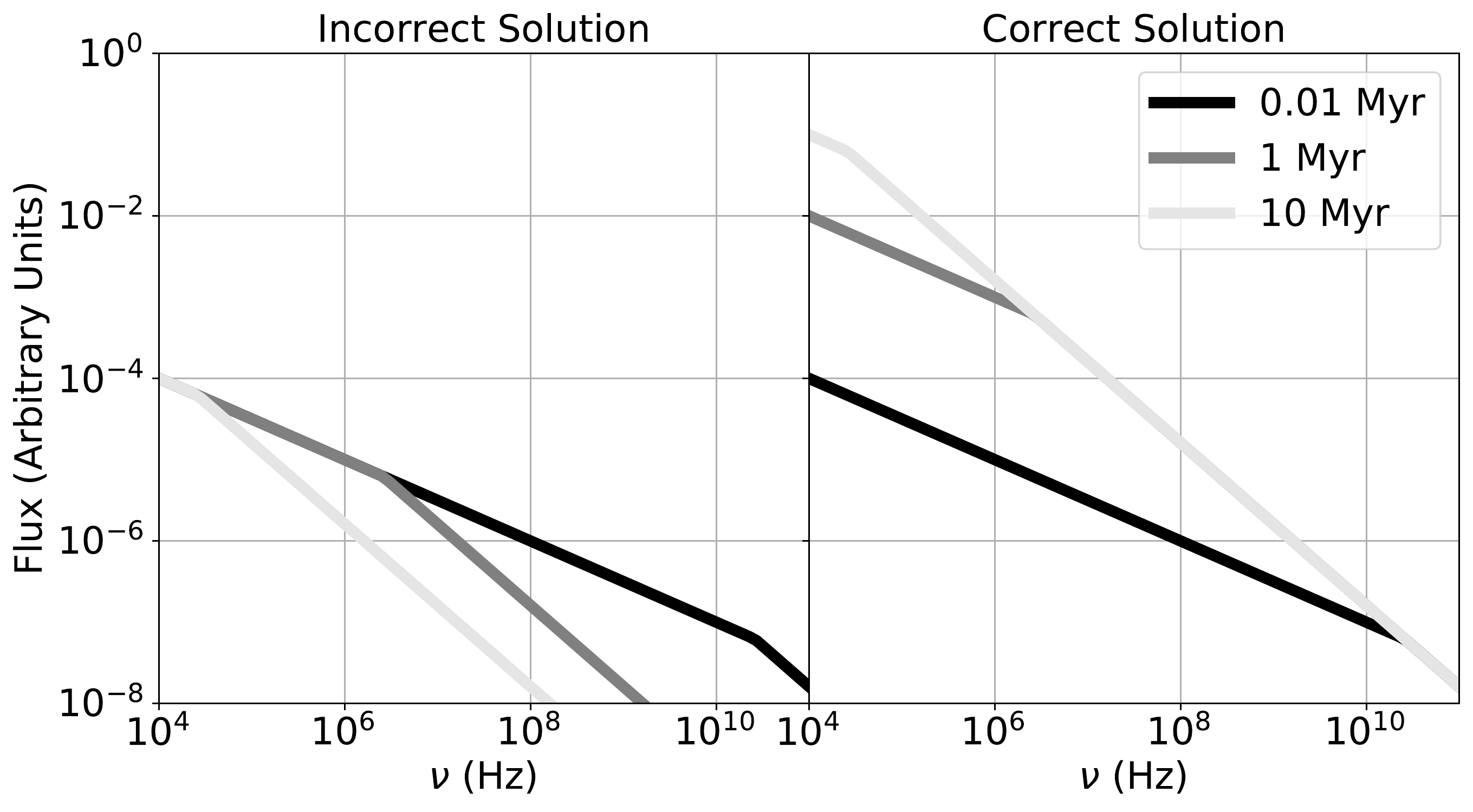}
    \caption{Left: The evolution of a spectrum posited by S18. As time progresses, a break in the specrum moves from blue to red with all fluxes blue of the spectral break falling as $\sim \nu^{-\alpha_R - 0.5}$ and all fluxes redward of $\nu_B$ remaining constant. Right: The correct time evolution for a spectrum with continuous injection (\citet{Kardashev:1962} equation 17) actually has fluxes increase for $\nu \lesssim \nu_B$ while fluxes with $\nu \gtrsim \nu_B$ remain constant in time.}
    \label{fig:Solutions}
\end{figure}

The basis for S18's second claim is that emission from sources powered by continuously injected particles following a power law with a spectral index of $\gamma$ experience spectral steepening blue-ward of a spectral break at $\nu_B \approx 2.6 \times 10^{-3} (B/\text{Gauss})^{-3} (t/\text{Myr})^{-2}$\,GHz so that $\alpha_R \approx (\gamma-1)/2$ for $\nu \lesssim \nu_B$ and $\alpha_R \approx \gamma/2$ for $\nu \gtrsim \nu_B$. S18 incorrectly assume that a source starts its life with a synchrotron spectral index of $(\gamma-1)/2$ across all frequencies and as time progresses, the break moves redward. Eventually $\nu_B$ falls below $1.4$\,GHz and after enough time, significantly reduces the amplitude of emission just blue of $21$\,cm which contributes the majority of absorbed emission. We illustrate this incorrect continuity solution in the left-hand panel of Fig.~\ref{fig:Solutions}. Given this scenario, S18 finds that 1.4\,GHz emission is reduced by a factor of $\sim 1000$ after 18\,Myr.

The correct solution to the continuity equation with continuous particle injection (e.g. \citealt{Kardashev:1962,Pacholczyk:1970}) is that the source starts out at $t=0$ with zero flux across all frequencies (and $\nu_B = \infty$). As time progresses, electrons pile up at low frequency below $\nu_B$ so that the overall amplitude of the spectrum rises with time as $\nu_B$ moves from high to low frequency (\citealt{Kardashev:1962}, equation 17). The shorter Synchrotron cooling time in a $\sim$\,mG emission region therefore leads to an excess of low-frequency emission at a fixed age rather then a dearth of high frequency emission (as supposed by S18). We illustrate the correct evolution of a constant B-field synchrotron emitting region with continuous particle injection in the right-hand panel of Fig.~\ref{fig:Solutions}. If S18's solution were true, then the prominent radio hot-spots that constitute a substantial fraction of emission from $\sim 10^7$\,year old FRII sources (where electron synchrotron lifetimes are $\sim 10^4$\,years)  would similarly be $\lesssim 1000$ times fainter then we observe them to be. The breaks in hot-spots actually remain at $\gtrsim 10$\,GHz since electrons advect away into the lobes after a short amount of time and continue to radiate in the lobe and cavity regions \citep{Meisenheimer:1989,Carilli:1991}. However, if these electrons were somehow kept within the hot-spot they would lower the hot-spot break frequency and simply increase flux significantly above what is is currently observed for $\nu<\nu_B$. It is possible that just like the electrons in the hot-spots of $z\approx1$ radio galaxies, electrons in $z \approx 17$ sources might quickly advect away from regions of large magnetic field strengths and bleed their remaining energy through IC scattering. Since the emitting electrons in such a source would only age for the time it takes for them to leave the emission region ($\lesssim 10^3$ years for pc - kpc scales), the spectral break could remain at $\gtrsim 10$\,GHz and not affect the frequencies relevant for 21\,cm absorption. For this reason, we considered two possibilities for spectral indices; $\alpha_\text{R} = 0.5$ (electrons are not trapped in the SE region) and $\alpha_\text{R} = 1.1$ (electrons are trapped in the SE region).

Having argued that continuously injected sources with sustained $\gtrsim 1$\,mG fields {\it can} potentially explain a $z \gtrsim 17$ background, {\it regardless of spectral aging}, we now discuss under what conditions such sustained magnetic fields might exist in the jet-impacted environments of intermediate mass black holes. For reference, 1\,mG is significantly higher then the equipartition fields typically detected in FRII radio lobes but it is comparable to the $\sim 100-1000$\,$\mu$G fields in hot spots \citep{Meisenheimer:1989}. Compact Steep Spectrum (CSS) and GHz-Peaked Spectrum (GPS) sources typically posses magnetic fields in the $1-10$\,mG range \citep{Murgia:1999,Murgia:2003}, significantly above the S18 threshold. Hence, the population of GPS and CSS sources are an example of synchrotron sources in the local Universe whose radio emission would survive at high redshift. GPS and CSS sources tend to have 2.7\,GHz luminosities of $\sim 10^{27}$\,W~Hz$^{-1}$ \citep{Odea:1998} and are powered by black holes with masses of $m_\bullet \sim 10^8 \text{M}_\odot$ \citep{Wu:2009}. The $z \approx 17$ black holes that we consider in this work have masses of $\sim 10^3 \text{M}_\odot$ and have radio luminosities at 2.7\,GHz of $\approx 10^{22}$W Hz$^{-1}$ which is sensible if one scales the radio luminosity down with black hole mass by five orders in magnitude. A source's minimum energy equipartition magnetic fields $(B_\text{eq})$ depend on both the radio luminosity of the source and the volume of the emitting region (e.g. \citealt{Wilson:2013}),
\begin{equation}\label{eq:BEQ}
    B_\text{eq} = \frac{1}{c} \left( \frac{3}{2} \frac{G}{H V} L_\nu \nu^{\alpha_\text{R}} \right)^{2/7},
\end{equation}
where $G$ and $H$ are functions of fundamental constants, $\alpha_\text{R}$ and the lower/upper frequency limits of SE (we assume typical values of $10$\,MHz and $100$\,GHz). If SE occurs in spherical region with radius $R_s$, we can solve for $R_s$ by rearranging~\ref{eq:BEQ}, 
\begin{equation}
    R_s = \left[ \frac{9}{8\pi} \frac{G}{H} \left(c B_\text{eq}\right)^{-7/2} L_\nu \nu^{\alpha_\text{R}} \right]^{1/3}.
\end{equation}
In the case of a positron-electron jet, a source with $L_\nu = 10^{22}$\,W Hz$^{-1}$ at $2.7$\,GHz must be contained within $R_s \approx 18$\,pc for minimum energy fields to support SE over IC scattering. Such an arrangement is plausible but faces the problem that it will quickly expand under the internal pressure of its constituent magnetic fields and relativistic particles. Indeed, most GPS/CSS sources are thought to evolve beyond their compact state after $\approx 10^6$ years \citep{Bicknell:1997,Murgia:2003}. We envision three potential scenarios in which SE from a particular black hole might be sustained over the 100\,Myr accretion lifetimes considered in this paper.  
\begin{enumerate}
    \item Ambient baryon loading. The ``old frustrated scenario" for explaining CSS and GPS sources is disfavored in most cases because insufficient baryons are present to contain the radio jet for $\sim 10^7$ years. Simulations by \citet{DeYoung:1993} show that $\sim 10^{11}$\,M$_\odot$ of gas are required to contain a GPS source $\sim 10^5$ times more powerful then the low mass black holes we consider; though cold dense clumps of gas can lower this value by a factor of $\sim 100$ \citep{Carvalho:1994,Carvalho:1998}. A back-of-the-envelope calculation indicates that containment by ambient baryons is highly unlikely. The mass, $m_b$, necessary to frustrate a synchrotron emitting region with $B_\text{eq}\sim1$\,mG such that it advances a distance $R_s$ through the ISM after some time  $T$ can be determined by setting the internal pressure of the synchrotron emitting plasma to the ram pressure of displaced gas. 
    \begin{equation}\label{eq:SOURCELIFETIME}
        \rho_b \left(\frac{R_s}{T} \right)^2 = \frac{3 m_b}{4 \pi R_s^3} \left(\frac{R_s}{T}\right)^2  = \frac{7}{3} \frac{B_\text{eq}^2}{2 \mu_0}
    \end{equation}
    So that 
    \begin{equation}\label{eq:MB}
        m_b \sim \frac{14 \pi }{9 \mu_0} R_s (T B_\text{eq})^2.
    \end{equation}
    Subsituting the values $R_s =20$\,pc and $B_\text{eq} = 1$\,mG, we find that $m_b \approx 10^9-10^{11}\text{M}_\odot$ is required for 10-100\,Myr containment. This value exceeds the mass of available baryons in a typical atomic cooling halo by several orders of magnitude. 
    \item Kinetic loading by infalling gas. Within the Bondi radius ($\sim 10^{-3}$\,pc) and outside of the accretion-disk radius, radio jets could encounter gas in-falling at a substantial fraction of the speed of light. It is possible that the radio jet could be frustrated by the high-velocity inflows that must exist to feed a black-hole at super-Eddington rates. Kinetic containment would involve a sub-parsec SE region trapped inside of the Bondi radius. 
    \item Periodic emission episodes. If the accretion onto the black-holes was sporadic, they might undergo recurring periods of emission where bright synchrotron regions are formed, shine for a relatively short time and rapidly cool at the end of each accretion episode or after the SE region expands to scales where it rapidly cools through IC scattering, allowing fresh accreting gas fill in the the jet cavities. In this way, periodic episodes of radio activity might be sustained over a significant fraction of a black-hole's $100$\,Myr accretion lifetime. 
\end{enumerate}
In summary, we disagree with S18's argument that sources must be $1000\times$ more radio loud then in the local Universe to have a large impact on the CD absorption feature. Many hot-spot features in the local Universe have persisted for $\gtrsim 10^6$\,years with significantly shorter electron cooling times ($\sim 10^4$\,years. In addition, $\sim 1$\,mG magnetic fields are a common feature of CSS and GPS sources. That said, keeping magnetic fields in SE regions from diffusing below $1\,$mG is a significant problem for the existence of RL sources at $z \approx 17$. We have suggested three possible solutions to this problem. A simple calculation indicates that the first; containment by ambient baryons, is unlikely. We leave detailed consideration of the feasibility of the other two scenarios for future work.

\subsection{Can Radio Loud Black Holes Explain EDGES?}\label{ssec:EDGES}
The 21\,cm absorption feature described in B18 has several unusual characteristics that are difficult to explain with models driven solely by stellar backgrounds. These features include 
\begin{enumerate}
\item A large depth of $\sim 500$\,mK 
\item A narrow width of $\Delta z \lesssim 10 $. 
\item Steep sides that climb $500$\,mK over $\Delta z \lesssim 2$
\item A flat bottom. 
\end{enumerate}
In Fig.~\ref{fig:Temperatures}, we compare our models to the $68\%$ and $95\%$ confidence regions derived from the MCMC fit of the raw EDGES data between $60-94$\,MHz published by B18\footnote{\url{http://loco.lab.asu.edu/EDGES/EDGES-data-release/}}. None of our models agrees well with the EDGES signal though our Fast and Massive Seeds models best reproduce the steepness and timing of the EDGES trough though they are still far less steep.

Guided by the intuition built in \S~\ref{sec:ModelingResults}, we obtain a better fitting model from our Small Halos scenario by reducing the Salpeter time to $\sim 18.2$\,Myr, delaying the trough by adding a $60$\,Myr delay between seed formation and vigorous accretion, and trading off the radio luminosity of each source for a greater number of sources by setting $g_\text{bol}=0.001$, $f_\text{L}=1$, and $\mu_R=2.1$. The adjustments to the source radio properties yield a similar radio gain to our fiducial model while better satisfying source count constraints by reducing the flux of each source. 

We finally make fine adjustments to better match the trough by increasing the Hydrogen column depth to $1.8 \times 10^{24}$\,cm$^{-2}$, reducing $T_{\text{vir}\bullet}^\text{min}$ to 1000\,K, raising $z_\text{min}^i$ to 21.5, and reducing $\alpha_X$ to 0.5. 

Given our highly uncertain knowledge of black hole accretion during the cosmic dawn, these new parameters are no more or less plausible then the set of assumptions that went into the models in \S~\ref{sec:ModelingResults} though they are closer to expected limits. The decrease in $\tau_s$ and $g_\text{bol}$ would be simultaneously accomplished by increasing the duty cycle to unity or raising the Eddington factor, $\lambda$, by a factor of two (as allowed by numerous super-Eddington accretion models \citep{Ichimaru:1977,Narayan:1998, Mineshige:2000, Volonteri:2015}), while lowering $k^{-1}_\text{bol}$ by a factor of three. Similar $k^{-1}_\text{bol}$ values are routinely measured in Type-I AGN \citep{Lusso:2010}). Lowering $T_{\text{vir}\bullet}^{\text{min}}$ to 1000\,K, is still within the range of seed halos masses that might survive baryon-dark matter velocity offsets but pushes towards lower limits \citep{stacy:2010, Grief:2011, Fialkov:2012}. The EDGEs models reduce the average radio loudness. Since radio emission faces significant obstacles, it might even be considered more plausible than our fiducial Pop-III model. Similarly, some time-delay between seed formation and rapid accretion is not unexpected due to feedback effects \citep{Alvarez:2009}. 

\begin{figure}
\includegraphics[width=.48\textwidth]{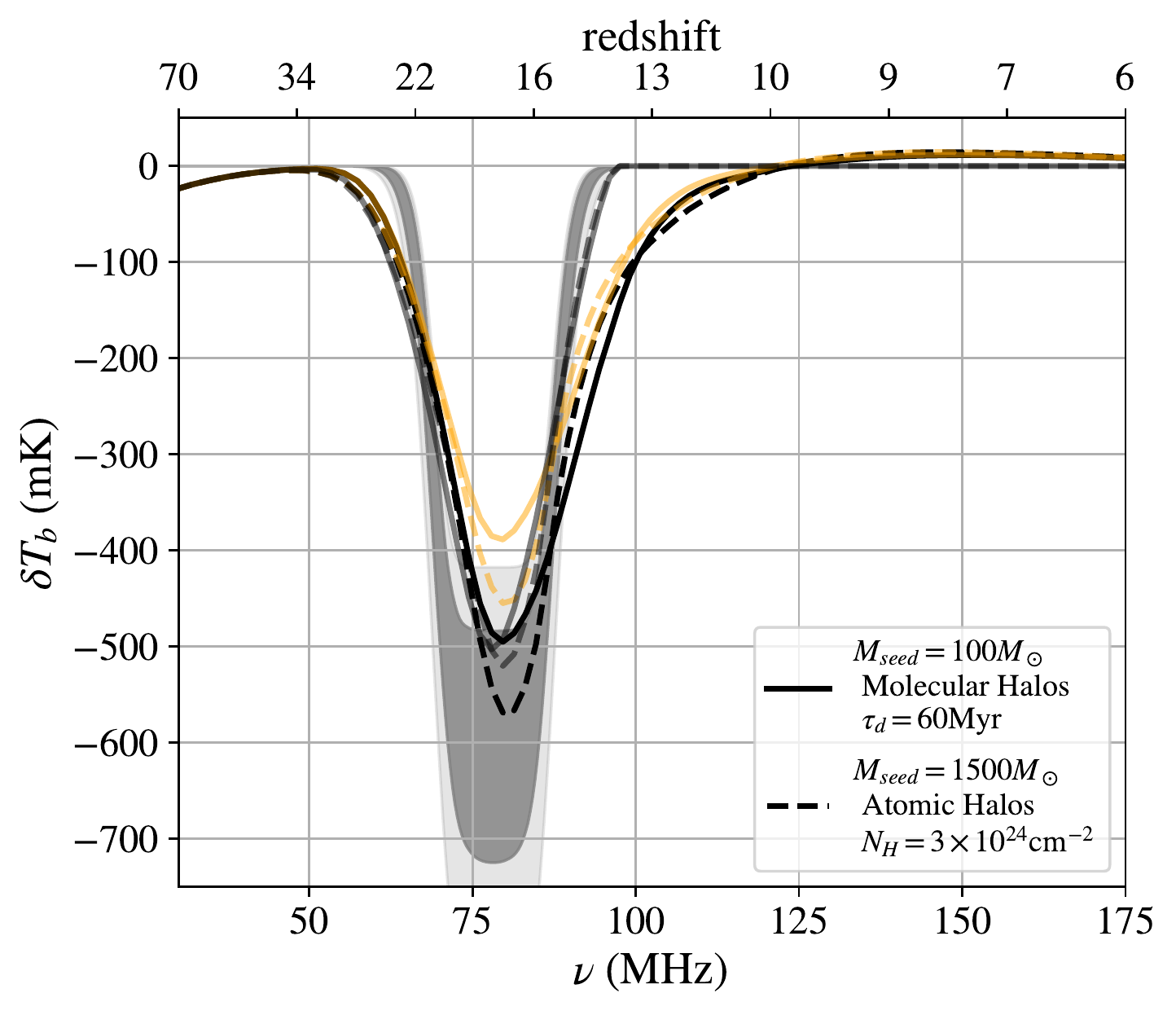}
\caption{$\delta T_b$ for two different EDGES-like models compared to the $68\%$ and $95\%$ contours of the detection in B18. We are able to produce the steep sides and large amplitude of the absorption trough through judicious choices of $\tau_\text{s}$, $T_{\text{vir}\bullet}^\text{min}$, and $\log_{10}N_\text{HI}$. However, a $\sim 100$\,mK absorption feature remains out to $z\approx 13$ unless significant HI ionization takes place by $z\approx 13$, inconsistent with recent Planck results (thin grey solid and dashed lines in this Figure and Fig.~\ref{fig:XHI_EDGES}). Within our modeling framework, we are not able to produce a flat-bottom trough.  
The orange light lines denote models where the radio spectral index ,$\alpha_R$ , has been flattened from 1.1 to $0.5$. Flatter spectral indices produce smaller absorption troughs as we found in Fig.~\ref{fig:Temperatures}.
}\label{fig:dTbEDGES}
\end{figure}

With these adjustments, we obtain a model that agrees roughly with the EDGES contours which we show in  Fig.~\ref{fig:dTbEDGES}. In Tables ~\ref{tab:TheOneTable} and~\ref{tab:EDGES}, we list these parameters that approximately fit 
the EDGES signal as EDGES Small Halos. 

\begin{figure*}
\includegraphics[width=\textwidth]{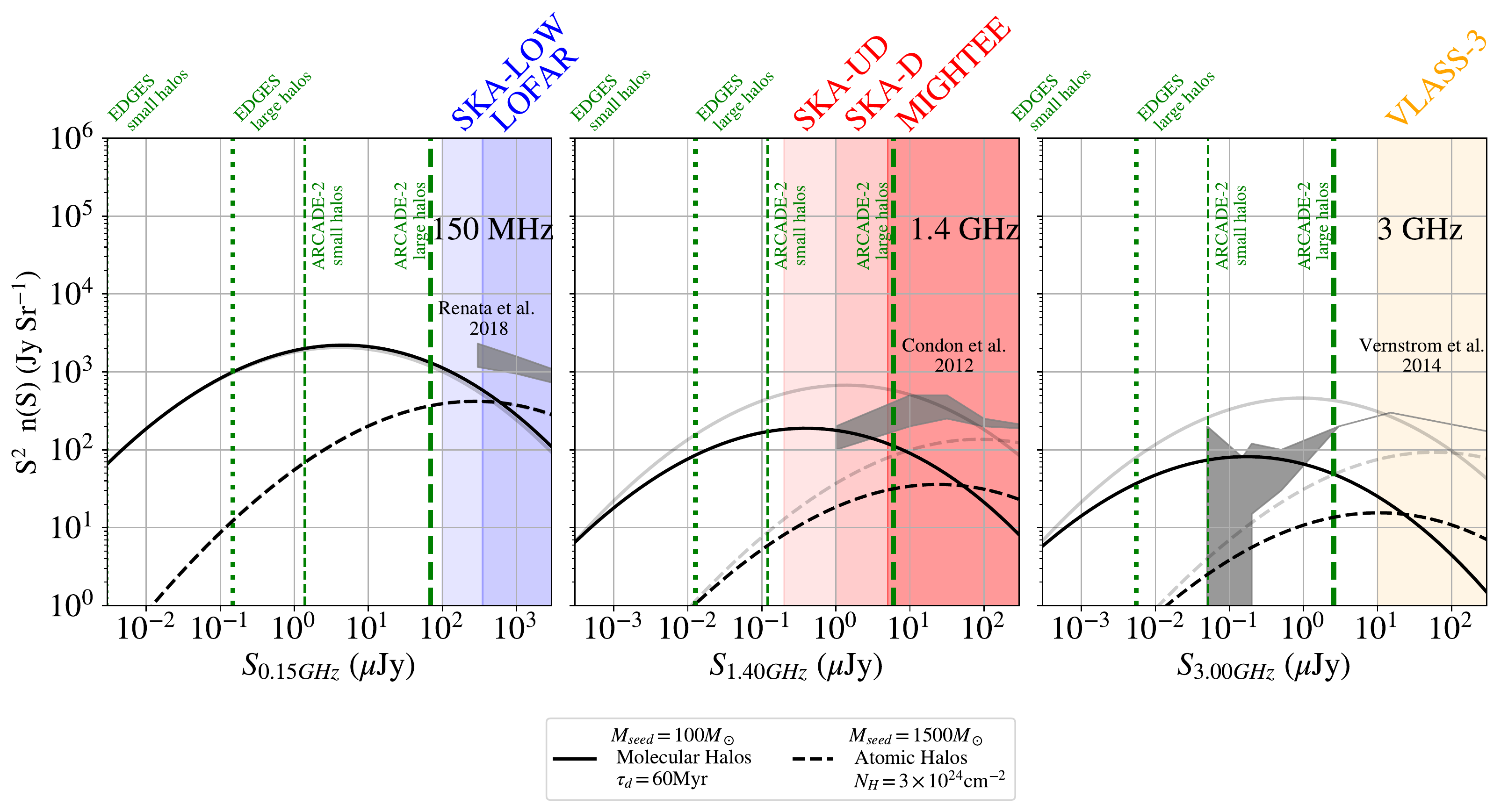}
\caption{The same as Fig.~\ref{fig:dndsEDGES} but now including the two EDGES-like models summarized in Table~\ref{tab:EDGES}. Both models are consistent with the results of recent fluctuation analyses. The distribution peaks of both the Small and Large Halos models are resolved by the deep and ultra-deep SKA1-MID surveys. Thus, these surveys will be able to place interesting constraints on models of radio-loud black holes that might explain the EDGES feature. Light lines indicate models where the radio spectral index, $\alpha_\text{R}$ has been flattened from 1.1 to 0.5. These flat spectral models are ruled out by existing fluctuation analyses.}
\label{fig:dndsEDGES}
\end{figure*}

Since it is difficult to obscure black holes in molecular cooling halos, we obtain a similar EDGES-like model with atomic cooling halos by starting with the EDGES Small Halos model, increasing the temperature range of seed halos to between $T_{\text{vir}\bullet}^\text{min} = 10^4$\,K and $T_{\text{vir}\bullet}^\text{max} = 5 \times 10^4$\,K, the seed mass to $1500$\,M$_\odot$, and the neutral column depth to $N_{\text{HI}\bullet} = 3 \times 10^{24}$\,cm$^{-2}$. These properties are consistent with the accretion properties simulated by \citep{Pacucci:2015}. Our large halos model involves fewer halos by making each individual black hole $\gtrsim 10 \times$ brighter than in the Small Halos scenario. In doing so, it begins to brush up against constraints at $S \gtrsim 1 \mu$Jy derived from confusion (Fig.~\ref{fig:dndsEDGES}). By increasing $N_{\text{HI}\bullet}$, we achieve the EDGES amplitude without violating confusion constraints. 

\begin{figure}
\includegraphics[width=.5\textwidth]{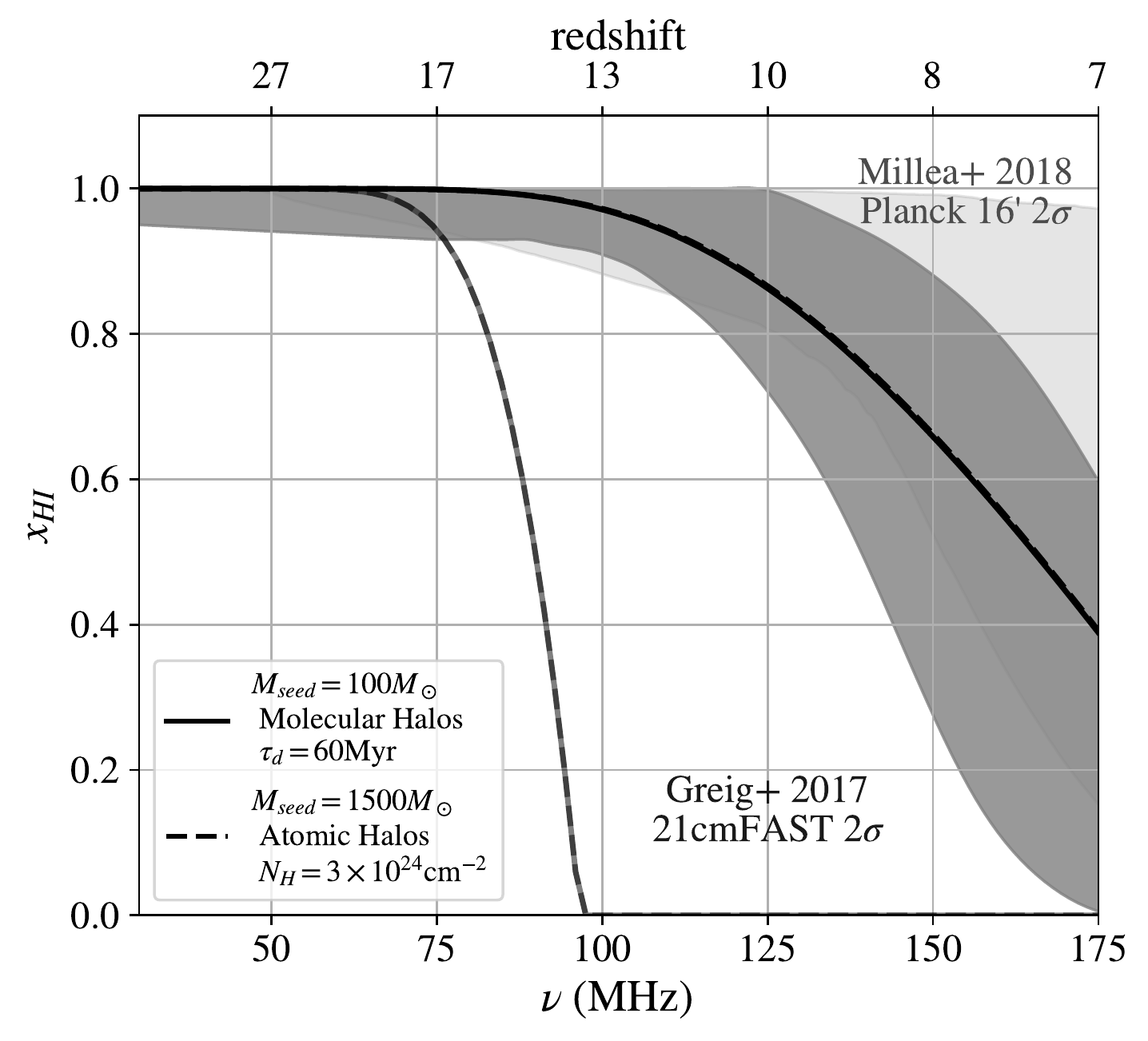}
\caption{The same as Fig.~\ref{fig:XHI} but now showing the two EDGES-like models summarized in Table~\ref{tab:EDGES}. Light sets of lines (which reionize earlier) represent scenarios identical to those in Table~\ref{tab:EDGES} where the stellar ionizing escape fraction has been raised to $f_{\text{esc}\star} = 1$ and $N_\gamma = 10^4$ in order to make the absorption feature more consistent with the B18 detection (Fig.~\ref{fig:dTbEDGES}). While increased stellar reionization brings better agreement with the B18 detection, it violates complementary constraints on reionization.}\label{fig:XHI_EDGES}
\end{figure}

We find two significant disagreements between our models and the EDGES signal that we are not able to mitigate through choosing ``better'' parameters. Firstly, $\delta T_b$ flattens out more gradually at $z\approx 14$, extending the end of the absorption feature beyond what the nominal EDGES detection. One way to reduce this flattening is to enhance the ionizating efficiency of SFGs by increasing $f_\text{esc}$ to unity and $N_{\gamma \star}$ to $10^4$. We show these fast ionization models as lightly colored lines in Fig.~\ref{fig:dTbEDGES}. Without raising the difficulties in achieving such a large ionizing flux, significant ionization at $z\approx 14$ would be in tension with other probes (Fig.~\ref{fig:XHI_EDGES}). These include measurements of Lyman Alpha Emitting galaxies (e.g. \citealt{McQuinn:2007}), the Ly-$\alpha$ damping wing in high redshift quasar spectra \citep{Mesinger:2004,Mesinger:2007}, the fraction of zero-flux pixels in Ly-$\alpha$ forest measurements \citep{Mesinger:2010}, constraints on $\tau_e$ from the CMB \citep{Planck:2016}, and kinetic Sunyaev Zeldovich (kSZ) measurements \citep{Trac:2011, Shaw:2012,Mesinger:2012}.  Secondly, none of our models appears to have a flattened bottom. 

\begin{table}
\begin{tabular}{l|l|l}
Parameter & EDGES small halos & EDGES large halos \\
\hline $T_{\text{vir}\bullet}^\text{min}$ & $1 \times 10^3$\,K & $1 \times 10^4$\,K \\
	   $T_{\text{vir}\bullet}^\text{max}$ & $1 \times 10^4$\,K & $5 \times 10^4$\,K \\
	   $m^i_\bullet$ & $100\,\text{M}_\odot$ & $1500\,\text{M}_\odot$ \\
       $\tau_\text{d}$ & $60$\,Myr & $0$\,Myr \\
       $\alpha_{\text{X} \bullet}$ & 0.5 & 0.5 \\
       $N_{\text{HI}\bullet}$ & $1.8 \times 10^{24}$ cm$^{-2}$ & $3 \times 10^{24}$ cm$^{-2}$ \\
       $\tau_\text{s}$ & 25\,Myr & 18\,Myr \\
       $z_{\text{min}}^i$ & 21.5 & 21.5\\
       $f_L$ & 1.0 & 1.0\\
       $\mu_\text{R}$ & 2.1 & 2.1\\
       $g_\text{bol}$ & 0.001 & 0.001
\end{tabular}
\caption{Parameter values for two models that are in approximate agreement with the EDGES detection (shown in Fig.~\ref{fig:dTbEDGES}). A smaller $\tau_\text{s}$ and larger $T_{\text{vir}\bullet}^\text{min}$ were required to match the location and steepness of the absorption trough. Larger X-ray column depths and lower radio-loudness fractions were introduced to obtain a large trough depth while staying below the limits imposed by the ARCADE-2 excess (see Fig.~\ref{fig:RadioBackgroundEDGES}). These parameters are also listed along with all other models in Table~\ref{tab:TheOneTable} }\label{tab:EDGES}
\end{table}

\begin{figure}
\includegraphics[width=.5\textwidth]{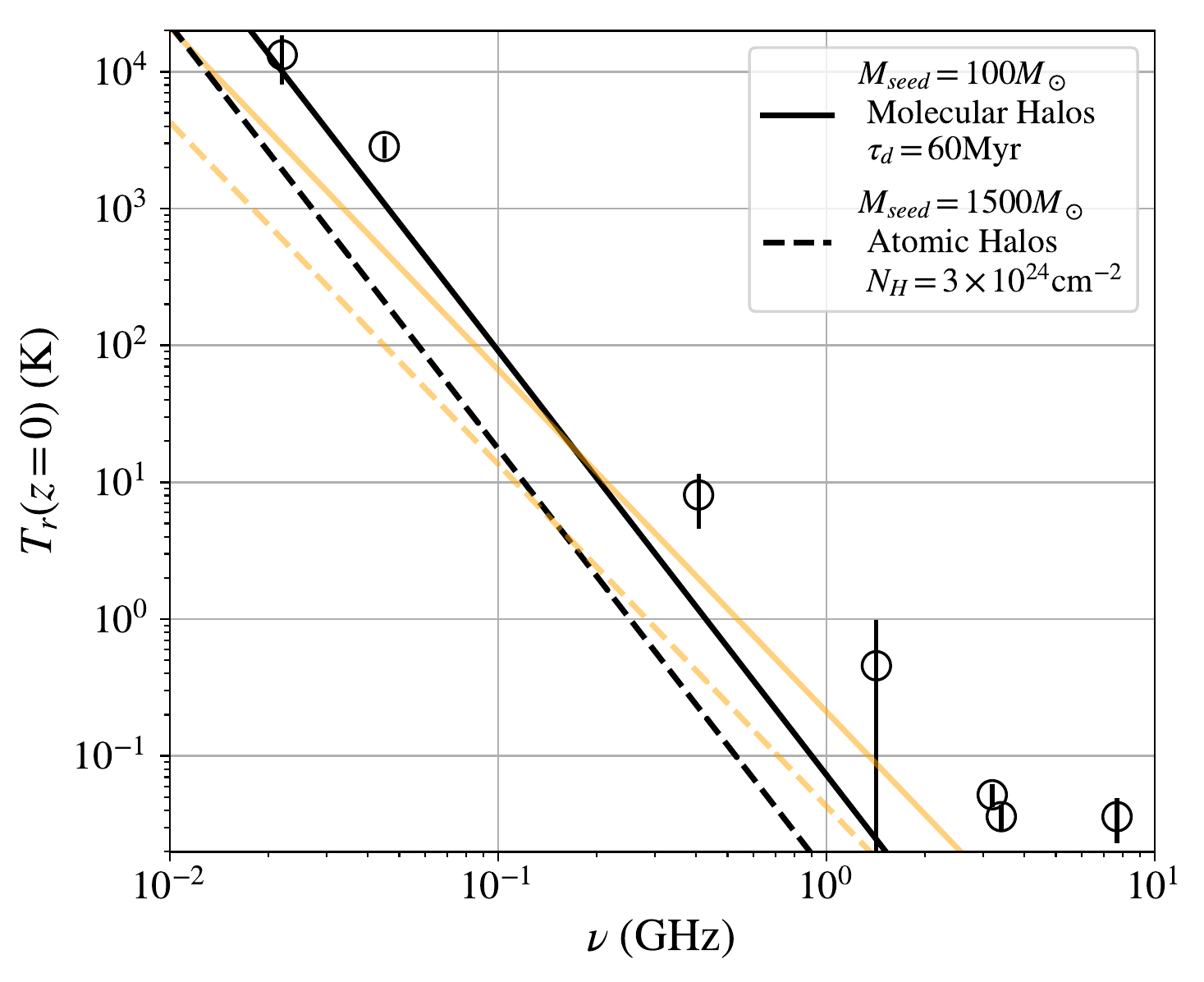}
\caption{The same as Fig.~\ref{fig:RadioBackground} except now only comparing radio backgrounds from our EDGES-like models (Table~\ref{tab:EDGES}) to other measurements of the excess radio monopole. Our EDGES-like models are below the level of excess. }\label{fig:RadioBackgroundEDGES}
\end{figure}

\begin{figure}
\includegraphics[width=.5\textwidth]{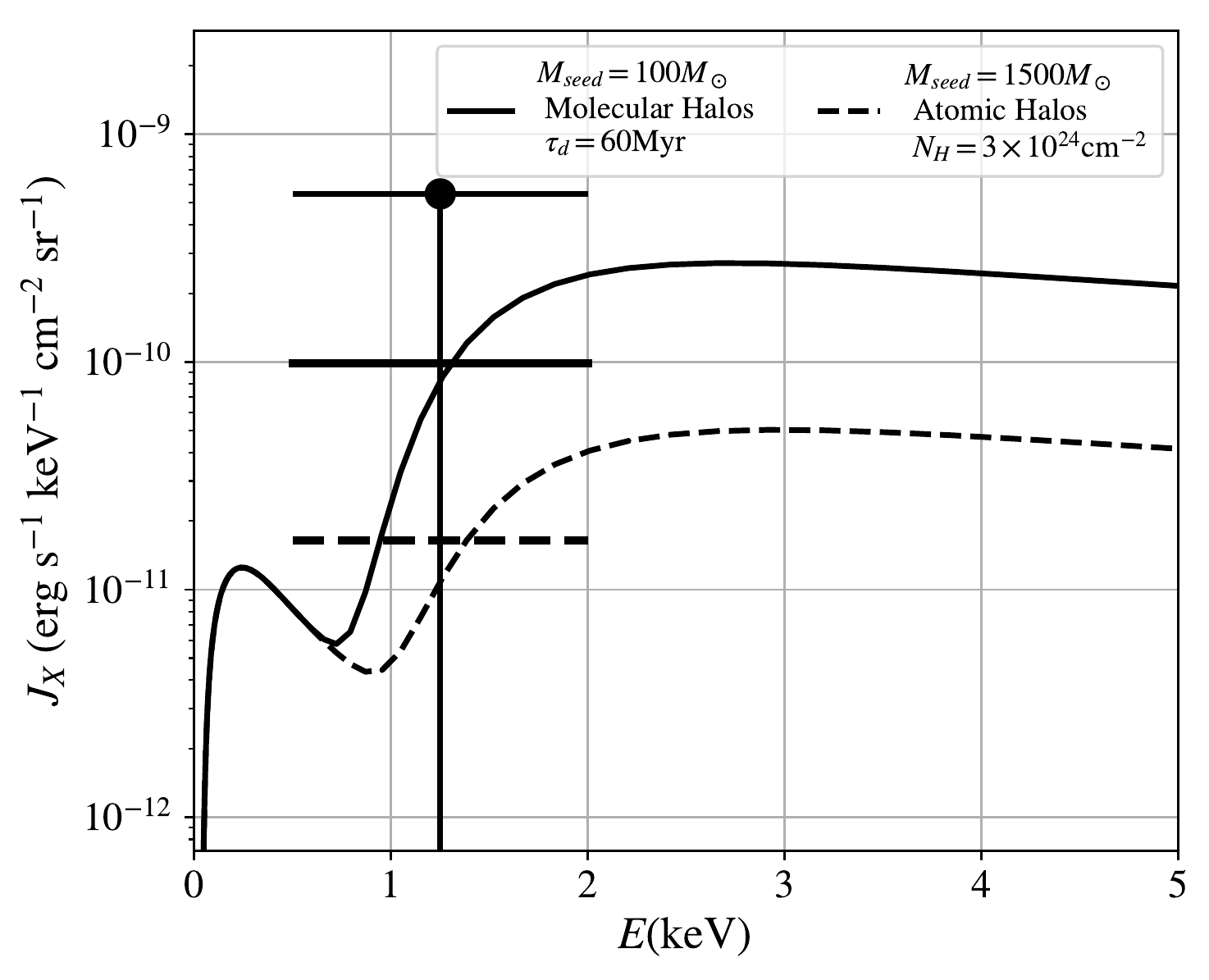}
\caption{The same as Fig.~\ref{fig:XRB} but now showing the X-ray background arising from the EDGES-like models in Table~\ref{tab:EDGES}.}\label{fig:XRBEDGES}
\end{figure}

We compare the X-ray and radio backgrounds from our EDGES-like models to existing limits in Figures~\ref{fig:XRBEDGES} and \ref{fig:RadioBackgroundEDGES}. We also show the source count distributions in Figs.~\ref{fig:dndsEDGES}. By construction, our Small Halos model lies just within existing radio-background constraints though the background amplitude might be made lower by increasing the obscuration of the black-holes. The XRB is similarly, thanks to obscuration, just below Chandra upper limits. Since our Large Halos model yields a population of $\gtrsim \mu$\,Jy sources that are constrained by confusion analyses, we do not have the freedom to produce such a strong radio background. Both the X-ray and radio backgrounds for the Large Halos model are well below that of Small Halos and limits in the literature.

To summarize, our model can explain the large depth and steepness of the EDGES feature. In order to be consistent with $\mu$Jy confusion analyses, models that involve halos above the atomic cooling limit are required to increase obscuration in order to preserve a large absorption amplitude and generate smaller fractions of the observed radio and X-ray backgrounds.

Our model is not able to reproduce the flat-bottom of the trough observed by B18 and in order to explain the rapid transition to $\delta T_b\approx 0$ at the end of the absorption feature, requires additional sources of ionizing photons (beyond what is provided by the obscured black holes and standard star formation scenarios). It is possible that these photons might be provided as black holes blow away their obscuring material, a process that we do not model in detail. However, an end to absorption feature consistent with EDGES and driven by ionization would contradict other probes. A more likely scenario might be enhanced heating from the same clearing. We have not attempted a systematic search of parameter space and are hesitant to do so until the specific properties of the EDGES feature are verified. Thus, it is entirely possible that other EDGES-like models do exist without the short-comings of our two examples. We leave more systematic fitting of the EDGES signal to future studies. 

We finally note that both of the scenarios that we have constructed to explain EDGES predict radio point-source populations that are above the detection threshold of future surveys on the SKA1-MID (Fig.~\ref{fig:dndsEDGES}). It follows that these future surveys have an important roll to play in validating or rejecting potential explanations of EDGES that involve discrete radio sources.

\section{Conclusions}\label{sec:Conclusion}

We have introduced a simple recipe for computing the impact of radio emission from black-hole seed growth on the 21\,cm signal that can be incorporated into existing semi-numerical simulation frameworks that estimate emissivities from the halo collapse rate (e.g. {\tt ares}, {\tt 21cmFAST}, and {\tt simfast21}). We combined our modeling framework with the 21\,cm global-signal formalism developed in \citet{Furlanetto:2006a}. For the first time, we compute the impact of radio emission from black holes self-consistently with the effects of their X-ray and UV emission. 

While we have focused on potential radio emission signatures of the first black holes, these same sources also generate X-ray emission and future deep X-ray surveys (e.g. from the proposed Lynx X-ray Observatory \citep{Gaskin:2019}) are also likely to provide additional constraints on the population of black-hole seeds at $z \gtrsim 15$. 

By studying the 21\,cm global signal arising in a handful of illustrative scenarios, we have arrived at the following conclusions on the signatures of radio-loud black-holes during the Cosmic Dawn. 

\begin{enumerate}
\item We find that if the first black-hole seeds were obscured by column depths $\gtrsim 10^{23}$\,cm$^{-2}$, radio-loudness levels similar to today will have a significant impact on the observed 21\,cm global signal -- at the level of tens to hundreds of percent. These significant effects are present even when the accretion rates are well below the Eddington limit. Thus, 21\,cm experiments should be able to constrain the radio-loudnesses of any rapidly growing black hole seeds that existed during and before the Cosmic Dawn absorption feature. This includes backgrounds from sub-nJy sources that are below the detection thresholds of any near term surveys. Hence, the global 21\,cm signal may be the best way to constrain the existence of such sources. That said, sub-Eddington accretion rates do not produce a signature as dramatic as EDGES and further work is needed to determine how radio signatures are degenerate with other model parameters.
\item When duty cycles and Eddington ratios are close to unity, obscured radio-loud black holes can generate a Cosmic Dawn absorption feature that is significantly deeper and narrower than what is seen in models where star formation dominates. This is due to the fact that the co-moving black-hole emissivity in our model has the freedom to evolve faster than the star-formation rate density. The steepness of the sides of the absorption feature increase with decreasing Salpeter time and/or increased seed mass while the timing is primarily affected by seed and halo mass. The overall depth is controlled by X-ray luminosity, spectral hardness, and obscuration. 
\item Black holes with radio-loudnesses similar to $z\approx1$ sources are able to explain many aspects of the reported EDGES absorption feature including its large depth, fast timing, and steep sides. While our model recreates the depth and steepness of the EDGES feature we were not able to produce a flat-bottomed absorption feature similar to what EDGES reports. In addition, our model cannot reproduce the rapid disappearance of absorption at the end of the trough unless it is achieved through reionization of the Intergalactic medium at $z \approx 14$, either through stellar contributions or the black-holes themselves (by clearing obscuring material). Unfortunately, both scenarios are inconsistent with CMB and Ly~$\alpha$ forest constraints.
\item A consequence of this model is that there should be a population of high-z radio sources that emerges at $\mu$\,Jy flux density levels but reproducing the EDGEs signal would not violate the ARCADE-2 limits. Current and future point source surveys have an important roll to play in better constraining the models presented in this work. The scenarios that we find to be most consistent with EDGES also predict populations of $\sim 1-10 \mu$Jy sources at 1.4\,GHz that greatly exceed expected counts from SFGs. In addtion, the seed mass/halo mass and accretion rate are all degenerate in the 21\,cm global signal. These degeneracies might be removed through direct detection of these sources by surveys since models with less abundant/more massive seeds yield individual sources with higher flux. We find that the nominal SKA1-MID deep survey described in P15 will be capable of resolving a proposed model with $1500$\,M$_\odot$ seeds in atomic cooling halos (Figs.~\ref{fig:dnds} and \ref{fig:dndsDelta}) which are accreting with a Salpeter time of $18$\,Myr while the ultra-deep survey can resolve sources in a scenario with $100$\,M$_\odot$ seeds in molecular cooling halos.

These conclusions are predicated on the uncertain assumption that black-holes can be radio-loud at $z\approx17$ in the first place. In \S~\ref{ssec:IC}, we found that the challenge of producing sustained radio emission boils down to containing SE within $\lesssim 20$\,pc regions over $\gtrsim$\,Myr time-scales or averaging over many episodic $ \lesssim 10^{5}$\,year emission episodes. We found that containment is a tall order for the limited ambient baryons present in primordial galaxies but it may also be achieved with accretion flows. We emphasize that one of the primary goals of our work was to demonstrate that 21\,cm and point source surveys have the potential to constrain such scenarios. Our last point raises the exciting prospect that decimeter-wavelength point source surveys and low frequency observations of redshifted 21\,cm may be used in conjunction to illuminate the unknown radio properties of SMBH progenitors.

\end{enumerate}

\section*{Acknowledgements}
We thank the anonymous referee, Michael Seiffert, Olivier Dor\'e, Jordan Mirocha, Luis Mas-Ruibas and Phillipe Berger for helpful comments on this manuscript. 
AEW's agknowledges support from the NASA Postdoctoral Program at the California Institute of Technology  Jet  Propulsion  Laboratory and the Berkeley Center for Cosmological Physics. Part  of  the  research  was  carried  out  at  the  Jet  Propulsion  Laboratory, California Institute of Technology, under a contract with  the  National  Aeronautics  and  Space  Administration.   

\section*{Code}
All code used in this project is publicly available at \url{https://github.com/anguta/global_bh}. The authors ackgnowledge extensive use of the {\tt colossus} library \citep{Diemer:2017} which can be downloaded at \url{https://bdiemer.bitbucket.io/colossus/}  along with the interpolation tables in {\tt 21cmFAST} \citep{Mesinger:2011} which is available at \url{https://github.com/andreimesinger/21cmFAST}. We also acknowledge use of the {\tt numpy} \citep{Numpy}, {\tt scipy} \citep{Scipy}, and {\tt matplotlib} \citep{Matplotlib} libraries. 




\bibliographystyle{mnras}
\bibliography{scream} 


\bsp	
\label{lastpage}
\end{document}